\documentclass{article}                    % onecolumn

\usepackage{graphicx}

\usepackage{latexsym}
\usepackage{amsmath}
\usepackage{amsfonts}
\usepackage{amssymb}
\begin{document}

\title{Detailed Examination of Transport Coefficients in Cubic-Plus-Quartic Oscillator Chains}


\author{G. R. Lee-Dadswell         \and
        B. G. Nickel \and
	C. G. Gray}


\maketitle

\begin{abstract}
We examine the thermal conductivity and bulk viscosity of a one-dimensional (1D) chain of particles with cubic-plus-quartic interparticle potentials and no on-site potentials.  This system is equivalent to the FPU-$\alpha \beta$ system in a subset of its parameter space.  We identify three distinct frequency regimes which we call the \emph{hydrodynamic regime}, the \emph{perturbative regime} and the \emph{collisionless regime}.  In the lowest frequency regime (the hydrodynamic regime) heat is transported ballistically by long wavelength sound modes.  The model that we use to describe this behaviour predicts that as $\omega \to 0$ the frequency dependent bulk viscosity, $\hat{\zeta} (\omega)$, and the frequency dependent thermal conductivity, $\tilde{\kappa} (\omega)$, should diverge with the same power law dependence on $\omega$.  Thus, we can define the \emph{bulk Prandtl number}, $Pr_{\zeta} = k_B \hat{\zeta}(\omega)/(m \hat{\kappa}(\omega))$, where $m$ is the particle mass and $k_B$ is Boltzmann's constant.  This dimensionless ratio should approach a constant value as $\omega \to 0$.  We use mode-coupling theory to predict the $\omega \to 0$ limit of $Pr_{\zeta}$.  Values of $Pr_{\zeta}$ obtained from simulations are in agreement with these predictions over a wide range of system parameters.   In the middle frequency regime, which we call the \emph{perturbative regime}, heat is transported by sound modes which are damped by four-phonon processes.  This regime is characterized by an intermediate-frequency plateau in the value of $\hat{\kappa}(\omega)$.  We find that the value of $\hat{\kappa}(\omega)$ in this plateau region is proportional to $T^{-2}$ where $T$ is the temperature; this is in agreement with the expected result of a four-phonon Boltzmann-Peierls equation calculation.  The Boltzmann-Peierls approach fails, however, to give a nonvanishing bulk viscosity for all FPU-$\alpha \beta$ chains.  We call the highest frequency regime the \emph{collisionless regime} since at these frequencies the observing times are much shorter than the characteristic relaxation times of phonons.
\end{abstract}

\newcommand{\rvec} 	{ {\mathbf{r}} 			}
\newcommand{\Rvec} 	{ {\mathbf{R}} 			}
\newcommand{\pvec}  { {\mathbf{p}} }
\newcommand{\jvec}  { {\mathbf{j}} }
\newcommand{\kvec} 	{ {\mathbf{k}} 			}
\newcommand{\qvec}  { {\mathbf{q}} }
\newcommand{\xvec}  { {\mathbf{x}} }
\newcommand{\vvec}  { {\mathbf{v}} }
\newcommand{\half}      {\frac{1}{2}}
\newcommand{\teth}     {\frac{3}{8}}
\newcommand{\quarter}    {\frac{1}{4}}
\newcommand{\partP}	    {\frac{\partial}{\partial P}}
\newcommand{\partB}	    {\frac{\partial}{\partial \beta}}
\newcommand{\where}	    {\noindent where }
\newcommand{\eqnk} {\stackrel{eq}{\rule{0cm}{0.1cm}}\!\! n_k}
\newcommand{\Xbar} {\overline{X}}
\newcommand{\xbar} {\overline{x}}
\newcommand{\infint} {\int_{-\infty}^\infty}

\section{Introduction}

The thermal transport properties of one-dimensional (1D) chains of particles have been recognized as an enigmatic puzzle for several decades \cite{Ford,Lepri_review}.  The problem is of more than academic interest because of the speculation that 1D systems may have applications for ``thermal management'' in nanoelectronics \cite{pap:Yu_etal_NanoLett5} and it has been suggested that the unusual properties of 1D systems might make a ``thermal transistor'' possible \cite{pap:thermal_transistor}.  By comparison, momentum transport in 1D chains has hardly been studied.  However, recently there has been work \cite{pap:mine_PRE72,Narayan_Ramaswamy} suggesting that thermal transport in 1D systems can be best understood by considering its coupling to momentum transport.  In \cite{pap:mine_PRE72} a theory is developed which allows the low frequency part of the heat current power spectrum to be predicted from the higher frequency parts of the heat current and momentum current power spectra.  The theory is shown to provide a good prediction for the special case of a system with the ratio of specific heats $\gamma = c_P/c_V =1$.  In such a system the heat current power spectrum can be predicted from the momentum current power spectrum alone.  One of the purposes of the present paper is to demonstrate that the theory developed in \cite{pap:mine_PRE72} also provides good predictions for the more general case of $\gamma \neq 1$.

Fourier's law of heat conduction is $\mathbf{J}_q = - \kappa \nabla T(\rvec, t)$ where $\mathbf{J}_q$ is the macroscopic heat flux density, $T(\rvec, t)$ is the local temperature and $\kappa$ is the thermal conductivity.  Newton's law of bulk viscous dissipation is $\mathbf{J}_p = - \zeta (\mathbf{\nabla} \cdot \mathbf{v}(\rvec, t))\hat{\mathbf{n}}$, where $\mathbf{J}_p$ is the macroscopic normal momentum current density (or stress) across a surface with normal direction $\hat{\mathbf{n}}$, $\zeta$ is the bulk viscosity and $\mathbf{v}(\rvec, t)$ is the local macroscopic velocity field.  We ignore shear viscosity since it is irrelevant for 1D systems.  The sense in which most 1D systems do not obey Fourier's law is that $\kappa$ fails to converge to a finite macroscopic value \cite{Lepri_review}.  Rather, $\kappa$ is seen to go as $N^{p}$, where $N$ is the number of particles in the chain and $p$ is some positive power.  The value of $p$ is a matter of great interest, with different values being reported for different systems \cite{pap:Grassberger_Nadler_Yang_PRL89,Lepri_review,Wang_etal}.  Some attempts have been made \cite{pap:mine_PRE72,Narayan_Ramaswamy} to theoretically predict the value of $p$.  While there are arguments in favour of universal behaviour \cite{pap:mine_PRE72,Lepri_review,Narayan_Ramaswamy}, the existence of universal behaviour is not supported by the bulk of simulation results that have been reported so far.  The consensus understanding of thermal conductivity in 1D systems can roughly be summarized as follows: the criteria for a system to have a finite conductivity are not known, for systems with infinite conductivity the dependence of the conductivity on system size is not understood and it is not known whether this dependence is universal in any sense.

Momentum transport in 1D systems is even more poorly understood, partly because it has rarely been discussed in the literature.  It has been predicted using mode coupling \cite{Ernst1,Ernst2} that the momentum current correlation function has a long time tail that goes as $t^{-1/2}$ in 1D \footnote{However, as was seen in \cite{pap:mine_PRE72}, unlike the case for the energy current correlation function, the amplitude of the $t^{-1/2}$ tail in the momentum current correlation function can vanish in some systems.}.  This is equivalent to a momentum current power spectrum going as $\omega^{-1/2}$ for small $\omega$ or a bulk viscosity that goes as $N^{1/2}$.  Given that the same calculations predict a $t^{-1/2}$ time tail for the heat current correlation function, which is not supported by any simulation results, this prediction should be viewed with scepticism.  

The failure of the mode-coupling calculation in 1D is not unexpected.  A key assumption is that there is a clear distinction between fast and slow relaxation processes and that for properties on long time-scales the effects of the fast processes can be incorporated into phenomenological hydrodynamic parameters.  This is roughly correct in 3D where phase space is dominated by large momenta and the multiplicity of the large momentum fast processes dominate.  In 1D the low momentum parts of phase space are much more significant and there are important relaxation processes on all time scales.  A separation into fast and slow is still qualitatively correct and perhaps even semi-quantitatively correct provided one generalizes the hydrodynamic parameters from fixed constants to time-scale dependent ones.

This picture was tested in \cite{pap:mine_PRE72} for the limiting situation in which mode-coupling theory predicted that, because of a vanishing thermodynamic amplitude, there are no slow momentum current relaxation processes.  Here we test this picture under more general circumstances.  If energy and momentum current correlations are dominated by the same fast relaxation processes then mode-coupling predicts that these two correlation functions are related by calculable thermodynamic quantities.  This is a refutable proposition.

We present simulation results for the FPU-$\alpha \beta$ model in which the interaction has been tuned from nearly harmonic through highly anharmonic and asymmetric to ultimately the highly anharmonic but symmetric pure quartic model studied in \cite{pap:mine_PRE72}.  The thermodynamic amplitudes predicted by mode-coupling theory that we test show a very non-trivial and non-monotonic variation and our simulations track this variation exceptionally well in the highly anharmonic and asymmetric regime and are not inconsistent at the two extremes.  As the harmonic limit is approached the comparison with simulation becomes difficult because all relaxation processes slow down.  At the other extreme, as the pure quartic model is approached the slow relaxation part of the momentum current correlations becomes too small to observe.

Because our simulation results confirm the predictions of mode-coupling in the limited sense described above, they also lend credence to the ``mode cascade'' toy model we proposed in \cite{pap:mine_PRE72}.  However, it is also clear that a complete validation of the toy model by simulation will be very difficult because of the enormous range of time scales that will have to be tracked.

The outline for the rest of the paper is as follows.  In the remainder of this introduction we provide some general background necessary for understanding the details of our simulations and comparisons with theory.  This is followed in section 2 by some specific mode-coupling background.  Sections 3 and 4 are a description of our model and the thermodynamic calculations we have had to perform.  Section 5 gives the numerical results of our simulations and is followed by a concluding section 6 which is a more extensive discussion than is given above.  Appendix A gives details on the symplectic algorithm that we are using in our molecular dynamics simulations.  Appendices B and C discuss issues not directly related to mode-coupling but for which our simulations have provided insight.  We emphasize throughout the paper that it is important to distinguish different frequency regimes by the processes that are important, but in practice this is not always easy to do as there is, as yet, little guidance from theory.  Appendix B is a comparison of Boltzmann-Peierls phonon scattering predictions for a weak coupling regime of the FPU-$\beta$ model.  The theoretical calculations are limited to the relaxation-time approximation so that, while the agreement with simulation is not perfect, it does suggest that Boltzmann-Peierls can be a valid description in a well defined parameter range for the thermal conductivity; however, as discussed in Appendix B, the Boltzmann-Peierls approach gives a vanishing result for the bulk viscosity.  Appendix C discusses the even simpler situation of no phonon scattering and is verified, for the momentum current correlations, to be reliable for the uppermost frequency range of the FPU-$\beta$ model.

The thermal conductivity, $\kappa$, and the bulk viscosity, $\zeta$, are formally related to the generalized or wave-vector and frequency dependent transport coefficients $\kappa (k, \omega)$ and $\zeta (k, \omega)$ by \cite{Forster}

\begin{subequations}
\label{eq:trans_coeff_defn}
\begin{eqnarray}
\kappa & = & \lim_{\omega \to 0} \hat{\kappa} (\omega) = \lim_{\omega \to 0} \lim_{k \to 0} \kappa (k,\omega) \\
\zeta & = & \lim_{\omega \to 0} \hat{\zeta} (\omega) = \lim_{\omega \to 0} \lim_{k \to 0} \zeta (k, \omega) \, .
\end{eqnarray}
\end{subequations}

\noindent Here and in the following we use the notation 

\begin{equation}
\label{eq:hat_defn}
\lim_{k \to 0} A(k) \equiv \hat{A} \, .  
\end{equation}

\noindent The frequency dependent $\hat{\kappa} (\omega)$ and $\hat{\zeta} (\omega)$ can be written as Green-Kubo relations in terms of the corresponding equilibrium heat current correlation function (HCCF), $\hat{C}_{\kappa} (t)$, and momentum current correlation function (MCCF), $\hat{C}_{\zeta} (t)$, namely

\begin{subequations}
\label{Kubo_relns}
\begin{eqnarray}
\hat{\kappa} (\omega) & = & \lim_{t \to \infty} \! \frac{\beta^2 k_{B}}{2}\!\! \int_{-t}^{t} \! \! dt^{\prime} e^ {i\omega t^{\prime}}\hat{C}_{\kappa}(t^{\prime}) \\
\hat{\zeta} (\omega) & = & \lim_{t \to \infty} \frac{\beta}{2} \! \int_{-t}^{t} \! \! dt^{\prime} e^{i\omega t^{\prime}} \hat{C}_{\zeta}(t^{\prime}) \, ,
\end{eqnarray}
\end{subequations}

\noindent where $k_B$ is Boltzmann's constant and $\beta = 1/k_B T$ is the inverse equilibrium temperature.  In terms of the corresponding currents $\hat{j}_{\mu} (t)$, where $\mu = \kappa$ or $\zeta$, we have   

\begin{equation}
\label{eq:corfncndefn}
\hat{C}_{\mu} (t) \equiv \lim_{L \to \infty} \frac{1}{L} \left \langle \delta \hat{j}_{\mu} (t) \delta \hat{j}_{\mu} (0) \right \rangle \, ,
\end{equation}

\noindent where $L$ is the system length, $\langle \cdots \rangle$ denotes a canonical average and $\delta \hat{j}_{\mu} (t)$ is the $k \to 0$ limit of the deviation of $\hat{j}_{\mu} (k,t)$ from its equilibrium value.  A further remark on the definition of $\hat{j}_{\mu} (t)$ is warranted.  Following (\ref{eq:hat_defn}), $\hat{j}_{\mu} (t)$ is the zero $k$ limit of the current density $j_{\mu} (k, t)$.  Under this definition the heat current density, $\hat{j}_{\kappa} (t)$, is equivalent to what is often referred to as the ``total heat flux'', $J_{\kappa} (t) \equiv \sum_{i=1}^N j_{\kappa} (q_i, t)$ ($q_i$ is the position of the $i^{\textrm{th}}$ particle), in studies where Green-Kubo relations are used to calculate thermal conductivities \cite{Lepri_review}.

The HCCF, $\hat{C}_{\kappa}$, in the Green-Kubo relation above can freely be exchanged with the \emph{energy} current correlation function, ECCF, $\hat{C}_{\epsilon}$ \cite{Forster} defined analogously to $\hat{C}_{\kappa}$ but with the heat flux density, $\hat{j}_{\kappa}$, replaced with the energy flux density, $\hat{j}_{\epsilon}$.  Numerically, it is more convenient to calculate the ECCF and this is what we do in the simulations reported in this paper.  In practice, we work with the Fourier transformed versions or power spectra of the correlation functions, 

\begin{equation}
\tilde{\hat{C}}_{\mu} (\omega) = \int_{-\infty}^{\infty} dt \exp(i \omega t) \hat{C}_{\mu} (t) \, .  
\end{equation}

\noindent We refer to these as the momentum current power spectrum (MCPS) and the energy current power spectrum (ECPS).

As shown below, in our system the particle spacing is arbitrary.  This, as well as computational convenience, makes it useful to work in a particle counting scheme (Lagrange picture) rather than a spatial coordinates scheme (Euler picture).  In the Euler picture (spatial coordinate scheme) hydrodynamic densities and currents are expressed as integrals over coordinates containing delta functions of the form $\delta (\rvec - \mathbf{q}_i)$ where $\rvec$ is the spatial coordinate being integrated over and $\mathbf{q}_i$ is the coordinate of the particle labeled $i$.  In contrast, the Lagrange picture allows these quantities to be expressed simply as sums over all particles.  In the Lagrange picture our spatial Fourier transform over a 1D chain is defined as

\begin{equation}
\delta j_{\mu} (k,t) = \frac{1}{\sqrt{N}} \sum_{s=1}^{N} \delta j_{\mu}^{s-\frac{1}{2}} (t) e^{iks} \, ,
\end{equation}

\noindent where $k = \{-(N-1)\pi/N, \ldots , -2\pi/N,0,2\pi/N, \ldots, \pi \}$ and $\delta j_{\mu}^{s-1/2}$ is the deviation from mean current between particles $s$ and $s\!\!-\!\! 1$.  The prefactor of $N^{-1/2}$ is to keep our current independent of system size as an aid to comparison between runs.  With this definition, $\langle \delta \hat{j}_{\mu} (t) \delta \hat{j}_{\mu} (0) \rangle$ is not proportional to $N$ and we should revise (\ref{eq:corfncndefn}) by the replacement $1/L \to N/L = 1/\ell$, where $\ell$ is the system length per particle.  However, because $\ell$ is arbitrary, as will be shown below, it is preferable to also define particle based currents (i.e. to work in the Lagrange picture instead of the Euler picture).  Thus, we use throughout

\begin{equation}
\label{eq:corfncndefn2}
\hat{C}_{\mu} (t) \equiv \lim_{N \to \infty} \left \langle \delta \hat{j}_{\mu} (t) \delta \hat{j}_{\mu} (0) \right \rangle \, ,
\end{equation}

\noindent instead of (\ref{eq:corfncndefn}).

To see how to write a current $j_{\mu}^{s-1/2}$ we examine the energy current.  Consider the Hamiltonian of the chain

\begin{equation}
E = \sum_{i=1}^N \left [ \frac{p_i^2}{2m} + \frac{1}{2} \sum_{j \neq i} \phi_{i, j} \right ] \, ,
\end{equation}

\noindent where $\phi_{i, j}$ is the potential energy due to the interaction between particles $i$ and $j$.  We are currently only interested in the case of nearest neighbour interacions so the sum over potential energies becomes $\sum_{i=1}^N \phi_{i, i-1}$.  Let us divide this up between particles so that $E = \sum_{i=1}^N E_i$ where $E_i$ is the energy of the $i^{\textrm{th}}$ particle.  We can choose to divide the energy up in several ways, leading to several different definitions of $E_i$.  We choose

\begin{equation}
\label{CompUsefulEDefn1}
E_i = \frac{p_i^2}{2m} + \frac{1}{2} \left [ \phi_{i, i-1} + \phi_{i+1, i} \right ] \, .
\end{equation}

\noindent We insist that $\phi_{i, i-1}$ depends only on the difference, $q = q_i - q_{i-1}$, between the positions $q_i$, $q_{i-1}$ so that $\partial \phi_{i, i-1}/\partial q_i = - \partial \phi_{i, i-1}/\partial q_{i-1}$.  Taking the time derivative of $E_i$ we obtain

\begin{displaymath}
\frac{d}{d t} E_i = \frac{p_i}{m} \frac{d p_i}{d t} + \frac{1}{2} \left [ (\dot{q}_i - \dot{q}_{i-1}) \phi_{i, i-1}^\prime + (\dot{q}_{i+1} - \dot{q}_i) \phi_{i+1,i}^\prime \right ] \, ,
\end{displaymath}

\noindent where $\phi_{i,j}^\prime (q) = \partial \phi_{i,j}/ \partial q$ and $\dot{q_i} = d q_i / d t \equiv v_i$.  From Hamilton's equations, $\dot{p}_i = - \phi_{i, i-1}^{\prime} + \phi_{i+1,i}^{\prime}$, so that we get

\begin{equation}
\label{SimpEidot}
\dot{E_i} = \frac{1}{2}\left [ (v_{i+1} + v_i)\phi_{i+1, i}^\prime - (v_i + v_{i-1})\phi_{i, i-1}^\prime \right ] \, .
\end{equation}

\noindent Taking the convention that a flow to the right (positive direction) is positive we can rewrite (\ref{SimpEidot}) as

\begin{equation}
\label{eq:j_from_Edot}
\dot{E_i} = - j_{\epsilon}^{i+1/2} + j_{\epsilon}^{i-1/2} \, ,
\end{equation}

\noindent where $j_{\epsilon}^{i - 1/2}$ is the energy flow (to the right) from particle $i-1$ to particle $i$ and $j_{\epsilon}^{i + 1/2}$ is the energy flow from particle $i$ to particle $i+1$.  Comparing this with (\ref{SimpEidot}) we can identify\label{pgdefn:j_epsilon}

\begin{equation}
\label{CompUsefulJDefn1}
j_{\epsilon}^{i+1/2} = - \frac{1}{2} (v_i + v_{i+1})\phi_{i+1,i}^\prime \equiv \frac{1}{2} (v_i + v_{i+1})\tau_{i+1,i} \, .
\end{equation}

\noindent where $\tau_{i+1,i}$ is \label{pgdefn:tau} the local stress (in 1D simply equal to the force of one particle on the other) between particles $i+1$ and $i$.  A form of the energy current which often appears in the literature (e.g. \cite{ProsenCampbell}) is

\begin{equation}
\label{CompUsefulJDefn2}
J_\epsilon^{i} = -\frac{1}{2} \left [ \phi_{i+1,i}^\prime + \phi_{i,i-1}^\prime \right ] v_i \, .
\end{equation}

\noindent This definition comes about if one divides up the energy between bonds instead of dividing it up between particles (i.e. (\ref{CompUsefulEDefn1}) is replaced by $E_i = (p_i^2 + p_{i+1}^2)/4m + \phi_{i+1, i}$).  It is easy to verify that this leads to the same definition of the total energy current and so it is equivalent.  A similar derivation to the one for (\ref{CompUsefulJDefn1}) leads to a Lagrange picture expression for the momentum current, 

\begin{equation}
\label{eq:CompUsefulJzeta}
j_{\zeta}^{i+1/2} = - \phi_{i+1,i}^\prime = \tau_{i+1,i} \, .
\end{equation}

Both (\ref{CompUsefulJDefn1}) and (\ref{eq:CompUsefulJzeta}) are defined in terms of divergences.  Hence, they are arbitrary up to an additive constant which, if chosen to give zero mean current, makes $j_\mu$ and $\delta j_\mu$ identical.  For the momentum current a logical choice for this additive constant is the pressure.

\section{The Bulk Prandtl Number}

It has been known since pioneering studies in the 1960s and 1970s \cite{pap:Alder_Wainwright_PRL18,pap:Alder_Wainwright_PRA4} that the current correlation functions (\ref{eq:corfncndefn}) generally have long-time tails that decay as some power of time.  Mode-coupling theory was developed to explain this phenomenon.  It invokes a physical picture in which each transport mode in the system is coupled to every other transport mode.  There are many different formalisms of mode coupling theory using many different sets of assumptions.  A good summary of the most widely used formalisms can be found in \cite{PomeauResibois}.  One of the most rigorous mode coupling theories is that of Ernst, et. al. \cite{Ernst1,Ernst2,Ernst3}.  In this theory the main starting assumptions are that the system of interest is in local equibrium and that local equilibrium is established and maintained by processes which are fast whereas the variations of the hydrodynamic currents and the couplings between them occur via processes which are slow.  The precise meanings of ``fast'' and ``slow'' are neither defined as an input to the theory, nor are the meanings provided by the theory.  Presumably there must be some minimum separation in time scales between the fast and slow time scales in order for the theory to be valid.

Of interest to the present study is the prediction, developed in \cite{Ernst2}, of the leading order terms of the momentum current correlation function and the heat current correlation function.  These are predicted for a system of general dimensionality, $d$.  For 1D the prediction of \cite{Ernst2} can be easily shown \cite{phd_mine} for long times to be

\begin{subequations}
\label{eqs:Ernst_corr_funcns}
\begin{eqnarray}
\label{eq:Czeta}
\hat{C}_{\zeta} (t) & \simeq & \left [ \frac{M_{+-}}{(\Gamma_s)^{1/2}} + \frac{M_{HH}}{(2 D_T)^{1/2}} \right ] \left ( \frac{1}{4\pi t} \right )^{1/2} \\
\label{eq:Ckappa}
\hat{C}_{\kappa} (t) & \simeq & \left [ \frac{K_{+-}}{(\Gamma_s)^{1/2}} \right ] \left ( \frac{1}{4\pi t} \right )^{1/2} \, ,
\end{eqnarray}
\end{subequations}

\where $D_T = m \kappa /\rho c_P$ is the thermal diffusivity, $\Gamma_s = (\gamma - 1)D_T + D_{\ell}$ is the sound damping constant, $\kappa$ is the thermal conductivity, $D_{\ell} = (4\eta/3 + \zeta)/\rho$ is the longitudinal diffusivity, $\eta$ is the shear viscosity (which vanishes for 1D) and $\zeta$ is the bulk viscosity.  Here  $m$ is the mass per particle, $\rho$ is the mass density, $\gamma = c_P/c_V$ is the ratio of the specific heats, $c_P$ is the specific heat capacity per particle at constant pressure and $c_V$ is the specific heat capacity per particle at constant volume.  The remaining constants in (\ref{eqs:Ernst_corr_funcns}) are the thermodynamic quantities

\begin{subequations}
\label{eqs:Ms_and_K}
\begin{eqnarray}
\label{eq:Mpm}
M_{+-} & = & \frac{1}{\beta^2} \left [ 1 - \frac{\gamma -1}{\alpha_P T} + \frac{\rho}{c} \left ( \frac{\partial c}{\partial \rho} \right )_s \right ]^2 \\
\label{eq:MHH}
M_{HH} & = & \frac{1}{2 \beta^2} (\gamma - 1)^2 \left [ 1 \!-\! \frac{1}{\alpha_P c_P} \! \left ( \frac{\partial c_P}{\partial T} \right )_P \!\!+\! \frac{1}{{\alpha_P}^2} \! \left ( \frac{\partial \alpha_P}{\partial T} \right )_P  \right ]^2 \\
\label{eq:Kpm}
K_{+-} & = & \frac{c^2}{\beta^2} \, ,
\end{eqnarray}
\end{subequations}

\where $c$ is the adiabatic speed of sound, $s$ is the entropy per particle and $\alpha_P = (1/L) (\partial L/ \partial T)_P$ is the thermal expansion coefficient.

The Green-Kubo integrands in (\ref{eqs:Ernst_corr_funcns}) have several important features that are worth noting.  First of all, their power law dependence on $t$ is of great interest; the predicted $t^{-1/2}$ behaviour (equivalent to power spectrum behaviour $\tilde{\hat{C}}_{i} (\omega) \sim \omega^{-1/2}$ at small $\omega$) is not in agreement with more recent theoretical \cite{pap:mine_PRE72,LepriLiviPoliti,Narayan_Ramaswamy} or numerical \cite{pap:mine_PRE72,Lepri_review} results.  Thus, it is tempting to dismiss the prediction as entirely incorrect for 1D.  However, as was shown in \cite{pap:mine_PRE72}, in 1D a careful examination of (\ref{eqs:Ernst_corr_funcns}) for the case of $\gamma = 1$ yields a prediction that any $\hat{C}_{\zeta} (t) \sim t^{-1/2}$ term is zero (raising the possibility of a finite bulk viscosity) and $\hat{C}_{\kappa} (t) \sim t^{-1/2}$, and simulations in that same paper confirm this prediction for a chain with quartic interparticle potentials.  Much more importantly, we showed in \cite{pap:mine_PRE72} that by interpreting $\Gamma_s$ as a frequency dependent phenomenological parameter $\Gamma_s (\omega)$ obtained from simulations of $\tilde{\hat{C}}_\zeta (\omega)$ at relatively high frequency, the $\tilde{\hat{C}}_\kappa (\omega)$ could be predicted and agreed with simulations at low frequency.  Thus, it is possible that (\ref{eqs:Ernst_corr_funcns}) is only ``minimally wrong''' by which we mean that (\ref{eq:Czeta}) and (\ref{eq:Ckappa}) are approximately correct if they are interpreted as relations between time dependent $\Gamma_s$ and $D_T$ for short times and $\hat{C}_\zeta$ and $\hat{C}_\kappa$ at long times.  We do not specify what constitutes short or long; rather (\ref{eqs:Ernst_corr_funcns}) is to be viewed as a pair of renormalization group equations in which every $\hat{C}_\zeta$ and $\hat{C}_\kappa$ at long times becomes the input for the $\Gamma_s$ and $D_T$ at the now short times for the next iteration.  The toy model introduced in the appendix of \cite{pap:mine_PRE72} was meant to illustrate this feature.  If $\tilde{\hat{C}}_{\zeta} (\omega)$ and $\tilde{\hat{C}}_{\kappa} (\omega)$ have the frequency dependence $\omega^{-p}$ for some range of small $\omega$ then over a well defined range of much smaller $\omega$ they will vary as $\omega^{-q}$ with $q = (1-p)/(2-p)$.  This cascade will eventually terminate at the fixed point $q=p=p^{\ast}=(3-\sqrt{5})/2 \approx 0.382$.  Note that (\ref{eqs:Ernst_corr_funcns}) is a special case of this renormalization group flow; an initial constant $\Gamma_s$ and $D_T$ implying $p=0$ give rise to $q=1/2$ which is the dependence $\tilde{\hat{C}}_\zeta$ and $\tilde{\hat{C}}_\kappa \propto \omega^{-1/2}$ or equivalently the long time tails $\propto t^{-1/2}$ in (\ref{eqs:Ernst_corr_funcns}).  If we now set $p = 1/2$ then the next frequency power becomes $q=1/3$ and this is the universal behaviour predicted in \cite{Narayan_Ramaswamy}.  However, we do not agree with the authors of \cite{Narayan_Ramaswamy} that there are no further renormalizations.  In our view the $\omega^{-1/3}$ dependence of the current power spectra also applies only over a limited frequency interval and will, at even lower frequency, be replaced by an $\omega^{-2/5}$ dependence and ultimately by $\omega^{-p^{\ast}}$.

The simulations in \cite{pap:mine_PRE72} tested a very small part of this picture as they relied only on the vanishing of $M_{+-}$ and $M_{HH}$ for $\gamma = 1$ and that $K_{+-} = c^2/\beta^2$.  The latter simply expresses the fact that heat is carried by weakly damped sound modes and does not require the full blown machinery of mode-coupling theory.  In the present paper we describe much more significant tests.  In particular, one of the implications of our renormalization group interpretation of (\ref{eqs:Ernst_corr_funcns}) is that, at long enough times, $\Gamma_s$ and $D_T$ will have the same frequency dependence.  In this case the ratio $\lim_{t \to \infty} \hat{C}_{\zeta}(t) / \hat{C}_{\kappa}(t) = \lim_{\omega \to 0} \tilde{\hat{C}}_{\zeta} (\omega) / \tilde{\hat{C}}_{\kappa} (\omega)$ should approach a constant at sufficiently low frequencies.  We speculate, given the assumptions in \cite{Ernst2,Ernst3} that ``sufficiently low frequencies'' means on time scales longer than those of the processes which establish and maintain local thermal equilibrium.  

This ratio of transport coefficients, which we expect to be constant in the thermodynamic and long time limit, is similar to the Prandtl number defined as $Pr = \nu/ D_T$, where $\nu=\eta/\rho$ is the kinematic shear viscosity.  However, our ratio is frequency dependent and involves the bulk viscosity.  As a convenient dimensionless ratio we define the \emph{bulk Prandtl number}

\begin{equation}
Pr_{\zeta} \equiv \lim_{\omega \to 0} Pr_{\zeta} (\omega) \equiv \lim_{\omega \to 0} \frac{\tilde{\hat{C}}_{\zeta} (\omega)}{m \beta \tilde{\hat{C}}_{\kappa} (\omega)} = \lim_{\omega \to 0} \frac{k_B}{m} \frac{\hat{\zeta}(\omega)}{\hat{\kappa}(\omega)} \, .
\end{equation}

\noindent From (\ref{eqs:Ernst_corr_funcns}) and (\ref{eqs:Ms_and_K}) we obtain

\begin{equation}
\label{eq:Pr_basic}
Pr_{\zeta} = \frac{\beta}{m c^2} \left [ M_{+-} + \left ( \frac{\Gamma_s}{2 D_T} \right )^{1/2} M_{HH}\right ] \, 
\end{equation}

\noindent which is an implicit equation for $Pr_\zeta$ since

\begin{equation}
\label{eq:Gam_DT_ratio}
\frac{\Gamma_s}{D_T} = (\gamma - 1) + \frac{c_P \zeta}{m \kappa} = (\gamma -1) + \frac{c_P}{k_B}Pr_\zeta \, .
\end{equation}

\noindent As a first approximation we observe that $c_P \zeta / (m \kappa)$ is normally quite small for FPU systems of the type studied here.  Assuming this we can write 

\begin{equation}
\label{eq:Pr_approx}
{Pr_{\zeta}}^{\textrm{approx}} = \frac{\beta}{m c^2} \left [ M_{+-} + \left ( \frac{\gamma - 1}{2} \right )^{1/2} M_{HH} \right ] \, .
\end{equation}

\noindent We can find the exact $Pr_{\zeta}$ explicitly by noting that (\ref{eq:Pr_basic}) and (\ref{eq:Gam_DT_ratio}) combined are a quadratic in $Pr_\zeta$ which can be solved to give

\begin{equation}
\label{eq:Pr_soln}
Pr_{\zeta} =  \overline{M}_{s} + \frac{c_P}{4k_B}{\overline{M}_{h}}^2 + \overline{M}_h \sqrt{\frac{c_P}{2k_B}\overline{M}_{s} + \frac{{c_P}^2}{16{k_B}^2}{\overline{M}_{h}}^2 + \frac{\gamma - 1}{2}}\, ,
\end{equation}

\where we must take the positive solution of the quadratic for consistency with (\ref{eq:Pr_approx}) in the appropriate limit and where for compactness we have written

\begin{subequations}
\begin{eqnarray}
\overline{M}_s & = & \frac{\beta}{mc^2} M_{+-} \, ,\\
\overline{M}_h & = & \frac{\beta}{mc^2} M_{HH} \, .
\end{eqnarray}
\end{subequations}

\noindent It must be stressed that this result should only be correct in a low frequency range, below the frequencies of processes which establish and maintain local thermodynamic equilibrium.  We will see in section~\ref{sec:cub-quart_chain}, over the range of parameters where this frequency regime is accessible to simulation, that the above provides a good prediction of the bulk Prandtl number in the system of interest.

\section{The Cubic-Plus-Quartic Chain}
\label{sec:cub-quart_chain}

We consider a 1D chain system with nearest neighbour forces governed by a cubic-plus-quartic interparticle potential.  The Hamiltonian is

\begin{equation}
\label{eq:Hamiltonian}
\mathcal{H} (\pvec, \qvec) = \sum_{i = 1}^N \left [ \frac{{p_i}^2}{2m} + \frac{\alpha}{3}(q_i - q_{i-1}- a)^3 + \frac{B}{4} (q_i - q_{i-1} -a )^4 \right ] \, ,
\end{equation}

\where $p_i = mdq_i/dt$ is the momentum of the $i^{\textrm{th}}$ particle, $q_i$ is the position of the $i^{\textrm{th}}$ particle, $m$ is the mass of each particle, $a$ is the equilibrium interparticle spacing which is a fixed, arbitrary, system parameter and $\alpha$ and $B$ are force constants.  We use periodic boundary conditions so that $q_0 = q_N - L$.  This is the familiar FPU-$\alpha \beta$ system with the harmonic force constant set to zero.  We use $B$ for the coefficient of the quartic term in the potential so as not to confuse it with the inverse temperature, $\beta = (k_B T)^{-1}$.  We also choose to work with interparticle distance as coordinates, rather than absolute particle position, so we define $X_i = q_i - q_{i-1} -a$.

There are a number of statistical mechanical quantities of interest to us in this work and these are most easily calculated in an isobaric ensemble.  The statistical weight of a configuration is proportional to the Boltzmann factor modified by a pressure term and $L = q_N - q_0 = Na + \sum X_i$ is allowed to take on all possible values so that the partition function takes the form \cite{book:TodaKuboSaito_1}

\begin{eqnarray}
\exp{(-\beta G)} & = & \frac{1}{h^N}\int d\pvec d\qvec \exp{[-\beta(\mathcal{H} + PL)]} \nonumber \\
\label{eq:disect_part_func}
& = & e^{-N\beta P a} \frac{1}{h^N} \left [ \int_{-\infty}^{\infty} dp \int_{-\infty}^{\infty} dX F (\beta, P) \right ]^N
\end{eqnarray}

\noindent where $d\pvec d\qvec$ denotes the phase space volume element for particles $i = 1 \ldots N$, $h$ is Planck's constant, $G = G(\beta, P, N)$ is the Gibbs free energy and $P$ is the pressure (same as average force in 1D).  We note that there is no need for a prefactor of $1/N!$ in the partition function because each particle is connected only to its neighbours so that particle ordering makes the particles distinguishable.  The $N$ statistically independent factors, $F(\beta, P)$ are

\begin{equation}
\label{eq:Boltzmann_factor}
F(\beta,\, P) = \exp{\left [ -\beta \left ( \frac{{p}^2}{2m} + \frac{\alpha}{3}{X}^3 + \frac{B}{4}{X}^4 + PX\right ) \right ]} \, .
\end{equation}

\noindent We will restrict our chain simulations to zero pressure but the formal expression (\ref{eq:Boltzmann_factor}) that includes $P \neq 0$ is convenient for deriving various thermodynamic quantities.  By defining characteristic length and time scales and using these to express the dynamical variables $X$ and $p$ in dimensionless form we can eliminate much of the apparent parameter dependence in (\ref{eq:Boltzmann_factor}).  We choose length and time units

\begin{equation}
\label{eq:length_time_units}
\ell_0 = (\beta B)^{-1/4} \, , \quad
t_0 = \left ( \frac{m^2 \beta}{B} \right )^{1/4} \, ,
\end{equation}

\noindent and write

\begin{equation}
X = \ell_0 x \, , \quad
p = \frac{m \ell_0}{t_0} v \, ,
\end{equation}

\noindent where now $x$ and $v$ are dimensionless.  Our Boltzmann factor in these new dimensionless variables is

\begin{equation}
\label{eq:Boltzmann_factor_dim_vars}
F(\beta, \, P) = \exp{\left ( -\frac{{v}^2}{2} - \alpha^{\ast} \frac{x^3}{3} - \frac{x^4}{4} - P^{\ast} x \right )} \, ,
\end{equation}

\noindent which shows that the parameter space of the model is two-dimensional with two dimensionless parameters defined by

\begin{eqnarray}
\label{eq:dim_parameter_P}
P^{\ast} & = & \beta P \ell_0 \\
\label{eq:dim_parameter_a}
\alpha^{\ast} & = & \frac{\alpha \beta^{1/4}}{B^{3/4}} \, .
\end{eqnarray}

\noindent We are currently interested in the zero pressure system.  This leaves us with a parameter space of interest that is one-dimensional and we can study the whole parameter space by fixing any two of $\alpha$, $\beta$, and $B$ and varying the third.  Because we have previously studied the pure quartic case \cite{pap:mine_PRE72} it is convenient for us to fix $\beta = 1$, $B = 1$ and vary $\alpha$.  As can be seen in (\ref{eq:dim_parameter_a}), varying $\alpha$ is equivalent to varying the temperature, $T$.  To summarize, our units are defined by $m=1$, $B=1$ and $\beta=1$.

The $\alpha^{\ast} = 0$ case is the pure quartic chain studied in \cite{pap:mine_PRE72} while in the limit of $\alpha^{\ast} \to \infty$ the system approaches the harmonic chain.  This can be seen by expanding the potential, V, about its minimum.  With reference to (\ref{eq:Hamiltonian}) we define

\begin{equation}
\label{eq:Vdefn}
V(X) = \frac{\alpha}{3}X^3 + \frac{B}{4}X^4 = \frac{1}{\beta} \left ( \frac{\alpha^{\ast}}{3} x^3 + \frac{1}{4} x^4 \right ) \, ,
\end{equation}

\noindent This has its minimum at $x_{\textrm{eq}} = -\alpha^{\ast}$.  Expanding around the minimum using $x = x_{\textrm{eq}} + \delta x$ we can write the dimensionless, scaled potential as 

\begin{equation}
\label{eq:Vprime}
\beta V(\delta x) = \textrm{const.} + \frac{{\alpha^{\ast}}^2}{2} \delta x^2 - \frac{2\alpha^{\ast}}{3} \delta x^3 + \frac{1}{4} \delta x^4 \, .
\end{equation}

\noindent If (\ref{eq:Vprime}) is used in the Boltzmann factor the harmonic term limits the magnitude of $\delta x$ to $O(1/\alpha^{\ast})$.  Thus, the cubic and quartic terms will be $O(1/{\alpha^{\ast}}^2)$ and $O(1/{\alpha^{\ast}}^4)$ respectively and, hence, negligible in the $\alpha^{\ast} \to \infty$ limit.  Between the extremes of $\alpha^{\ast} = 0$ and $\alpha^{\ast} \to \infty$ the potential is both highly anharmonic and asymmetric and this is precisely what we need for a significant test of the Ernst formulae (\ref{eqs:Ernst_corr_funcns}) in a case distinct from that examined in \cite{pap:mine_PRE72}.  The fact that by varying $\alpha^{\ast}$ we can also explore the crossover from strong to weak anharmonicity is an added bonus.  Given the above discussion, removing the constraint of zero harmonic force constant in (\ref{eq:Hamiltonian}) would probably not yield much additional information.

We are now ready to discuss the various quantities involved in calculating $Pr_{\zeta}$: $c$, $c_P$, $\gamma$, $D_T$, $\Gamma_s$, $\alpha_P$, $\partial \alpha_P/\partial T$, $\partial c_P /\partial T$, $\partial c/\partial \rho$.  We will make use of the constant pressure partition function (\ref{eq:disect_part_func}) which reduces to

\begin{equation}
\exp{(-\beta G)} = \left ( \frac{2\pi m}{\beta h^2} \right )^{N/2} e^{-N\beta Pa} \left [ \int_{-\infty}^{\infty} dX \exp{\left [-\beta \left (V(X) + PX  \right )  \right ]} \right ]^N \, .
\end{equation}

\noindent with $V(X)$ given by (\ref{eq:Vdefn}).  In deriving the thermodynamic quantities of interest we repeatedly encounter averages of the form

\begin{equation}
\label{eq:Xn_defn}
\overline{X}_{n} (\beta, P) \equiv {\ell_0}^n \overline{x}_n = \langle {X}^n \rangle = \frac{\int_{-\infty}^{\infty} dX \, X^n \exp{\left [-\beta \left (V(X) + PX  \right )  \right ]}}{\infint dX \exp{\left [-\beta \left (V(X) + PX  \right )  \right ]}} \, ,
\end{equation}

\noindent which is the $n^{\textrm{th}}$ moment of $X$.  These moments satisfy the recursion relation

\begin{equation}
(n+1) \overline{X}_n = \beta \left [ \alpha \overline{X}_{n+3} + B \overline{X}_{n+4} + P \overline{X}_{n+1} \right ] \, .
\end{equation}

\noindent Hence, besides the trivial $\overline{X}_0 = \overline{x}_0 = 1$, we need only evaluate $\overline{X}_1$ and $\overline{X}_2$ as numerical integrals.  Furthermore, the temperature and pressure derivatives of $\overline{X}_n$ follow directly from (\ref{eq:Xn_defn}).  We get

\begin{subequations}
\begin{eqnarray}
\label{eq:XbyP} \partP \overline{X}_n & = & -\beta \left ( \overline{X}_{n+1} - \overline{X}_{n} \overline{X}_{1} \right ) \, , \\
\label{eq:XbyB} \partB \overline{X}_n & = & \frac{\alpha}{3} (\overline{X}_n \overline{X}_3 - \Xbar_{n+3}) \nonumber \\
& & + \frac{B}{4} (\Xbar_n \Xbar_4 - \Xbar_{n+4}) + P (\Xbar_n \Xbar_1 - \Xbar_{n+1}) \, .
\end{eqnarray}
\end{subequations}

The relations (\ref{eq:XbyP}) and (\ref{eq:XbyB}) form the basis for our subsequent derivation of the thermodynamic quantities.  For future reference we list below a number of the most important thermodynamic expressions characterizing the cubic-plus-quartic chain.  As a useful starting point, we obtain an equation of state via

\begin{equation}
\langle L \rangle = \frac{\partial G}{\partial P} = N(a + \Xbar_1 (\beta, P)) \, .
\end{equation}

\noindent and hence our equilibrium spacing between neighbouring particles is

\begin{equation}
\label{eq:eq_particle_spacing}
\ell = \frac{\langle L \rangle}{N} = a + \Xbar_1 = a + \ell_0 \xbar_1 \, .  
\end{equation}

\noindent The fact that (\ref{eq:eq_particle_spacing}) depends on the arbitrary spacing, $a$, implies that $\ell$ cannot appear in any expression for a thermodynamic quantity except as a trivial multiplier.  The average energy per particle is found from the enthalpy, $\langle E \rangle +P\langle L \rangle = \partial (\beta G)/\partial \beta$, and is

\begin{eqnarray}
\frac{\langle E \rangle}{N} & = & \left [ \frac{1}{2\beta} + \frac{\alpha}{3} \Xbar_3 + \frac{B}{4} \Xbar_4 \right ] \nonumber \\
& = & \frac{1}{\beta} \left [ \frac{1}{2} + \frac{\alpha^{\ast}}{3} \xbar_3 + \frac{1}{4} \xbar_4 \right ] \, .
\end{eqnarray}

\noindent We obtain the thermal expansion coefficient by differentiation of (\ref{eq:eq_particle_spacing})

\begin{eqnarray}
\label{eq:alpha_P}
\alpha_P & \equiv & -\frac{1}{\ell} \frac{1}{k_B T^2} \left ( \frac{\partial \ell }{\partial \beta } \right )_{P} \nonumber \\
& = & \frac{1}{\ell} \frac{1}{k_B T^2} \left [ \frac{\alpha}{3} (\Xbar_4 - \Xbar_3 \Xbar_1) + \frac{B}{4} (\Xbar_5 - \Xbar_4 \Xbar_1) \right . \nonumber \\
& & \left . + P (\Xbar_2 - {\Xbar_1}^2) \right ] \nonumber \\
& = & \frac{\ell_0}{T \ell} \left [ \frac{\alpha^{\ast}}{3} (\xbar_4 - \xbar_3 \xbar_1) + \frac{1}{4} (\xbar_5 - \xbar_4 \xbar_1) + P^{\ast} (\xbar_2 - {\xbar_1}^2) \right ] \, .
\end{eqnarray}

Similarly, the isothermal compressibility is

\begin{equation}
\label{eq:chi_T}
\chi_T =  - \frac{1}{\ell} \partP \ell = \frac{\beta}{\ell} (\Xbar_2 - {\Xbar_1}^2) = \beta \frac{{\ell_0}^2}{\ell} \left (\xbar_2 - {\xbar_1}^2 \right )\, .
\end{equation}

The specific heat capacity per particle at constant pressure is

\begin{eqnarray}
c_P & \equiv & \frac{C_P}{N}  = \frac{-k_B \beta^2}{N} \partB [ \langle E \rangle +P \langle L \rangle ] \nonumber \\
& = & k_B \left [ \frac{1}{2} + \frac{{\alpha^{\ast}}^2}{9} \left ( \xbar_6-{\xbar_3}^2 \right ) + \frac{\alpha^{\ast}}{6} \left (\xbar_7 - \xbar_4 \xbar_3  \right ) + \frac{1}{16} \left (\xbar_8 - {\xbar_4}^2 \right ) \right . \nonumber \\
& & + \frac{2P^{\ast}\alpha^{\ast}}{3} \left (\xbar_4 - \xbar_3 \xbar_1 \right ) + \frac{P^{\ast}}{2} \left (\xbar_5 - \xbar_4 \xbar_1  \right ) \nonumber \\
& & \left . + {P^{\ast}}^2 \left (\xbar_2 - {\xbar_1}^2  \right ) \right ] \, ,
\end{eqnarray}

\noindent and for the specific heat at constant volume one can use the identity $c_V = c_P - \ell T{\alpha_P}^2 /\chi_T$ with the results from (\ref{eq:alpha_P}) and (\ref{eq:chi_T}).  The specific heat ratio $\gamma = c_P/c_V$, combined with (\ref{eq:chi_T}) gives the adiabatic speed of sound $c = (1/\rho \chi_s)^{1/2}$, where $\chi_s = \chi_T / \gamma$ is the adiabatic compressibility, as 

\begin{equation}
c^2 = \frac{\ell \gamma}{m \chi_T} = \left (\frac{\ell}{t_0} \right )^2 \frac{\gamma}{\xbar_2 - {\xbar_1}^2} \, .
\end{equation}

As noted above, the sound speed varies linearly with $\ell$ as expected; this is its only dependence on the arbitrary spacing, $a$.  In fact the arbitrariness is absent in the Lagrangian picture which we adopt for our simulations; here the natural unit is $c/\ell$ and the speed of sound is viewed as being measured in terms of particle number per unit time (i.e. it is a ``hopping'' frequency for a disturbance to move from one particle to the next).

At zero pressure for both limits $\alpha^{\ast} = 0$ and $\alpha^{\ast} \to \infty$, the specific heat ratio $\gamma = 1$.  Numerical evaluation of the formulae above shows the maximum $\gamma \simeq 1.544$ at $\alpha^{\ast} \simeq 2.418$.  This value of $\alpha^{\ast}$ serves as a reasonable central value for our chain simulations.

\section{Evaluation of the Bulk Prandtl Number}

We wish to predict $Pr_\zeta$ as a function of $\alpha^{\ast}$.  In order to do this we need the thermodynamic expressions from the previous section.  We must also be able to evaluate the Ernst current amplitudes (\ref{eqs:Ms_and_K}).  Let us now reduce the expressions of the current amplitudes to a form which is useful for numerical evaluation for the cubic-plus-quartic model.  $K_{+-}$ is already in a convenient form.  We note that $\sqrt{K_{+-}}$ is clearly proportional to the mean lattice spacing, $\ell$, as is expected of a quantity proportional to a current.  But this same linear dependence is not evident in $M_{+-}$ and $M_{HH}$.  We rewrite them as

\begin{subequations}
\label{eq:M_reduction}
\begin{eqnarray}
M_{+-} & = & \frac{1}{\beta^2} {m_{+-}}^2, \quad m_{+-} \equiv \frac{\gamma - 1}{\alpha_P T} - \frac{\rho}{c} \left (\frac{\partial c}{\partial \rho} \right )_s - 1 \\
M_{HH} & = & \frac{1}{2\beta^2} {m_{HH}}^2, \quad m_{HH} \equiv (\gamma - 1) \left [ 1-\frac{1}{\alpha_P c_P} \left( \frac{\partial c_P}{\partial T} \right )_P \right . \nonumber \\
& & \left . + \frac{1}{{\alpha_P}^2} \left ( \frac{\partial \alpha_P}{\partial T} \right )_P \right ] \, .
\end{eqnarray}
\end{subequations}

\noindent We now express $m_{HH}$ and $m_{+-}$ in terms of the quantities $c/\ell$ and $\ell \alpha_P$ which do not depend on the arbitrary spacing parameter, $a$.  Doing this, and noting the structures of (\ref{eq:M_reduction}), which involves expressions of the form $(1/A)(\partial A/ \partial T)$ and $(1/A) (\partial A/\partial \rho)$ we are able to express them as

\begin{subequations}
\label{eq:m_by_ell}
\begin{eqnarray}
\label{eq:mpm_by_ell}
\frac{m_{+-}}{\ell} & = & \frac{\gamma -1}{\ell \alpha_P T} - \left ( \frac{\partial \ln{(\ell/c) }}{\partial \ell} \right )_s \, , \\
\label{eq:mHH_by_ell}
\frac{m_{HH}}{\ell} & = & \frac{\gamma -1}{\ell \alpha_P}\left ( \frac{\partial \ln{(\ell \alpha_P /c_P) }}{\partial T} \right )_P \, .
\end{eqnarray}
\end{subequations} 

\noindent The specific heat relation $c_P - c_V = \ell T {\alpha_P}^2 / \chi_T$ or $(\gamma - 1)/ (\ell \alpha_P) = \alpha_P \gamma T/(\chi_T c_P)$ allows us to eliminate the potentially vanishing denominators in (\ref{eq:m_by_ell}), so that these relations remain valid in the limits where $\gamma \to 1$ and $\alpha_P \to 0$.  We replace (\ref{eq:mHH_by_ell}) by

\begin{equation}
\frac{m_{HH}}{\ell} = \frac{\alpha_P \gamma T}{\chi_T c_P} \left ( \frac{\partial \ln{(\ell \alpha_P /c_P) }}{\partial T} \right )_P
\end{equation}

\noindent and note that this can be evaluated by the procedures described in section~\ref{sec:cub-quart_chain}.  These procedures do not apply directly to the derivative at constant $s$ in (\ref{eq:mpm_by_ell}) but we can start with the chain rule and write

\begin{equation}
\label{eq:chain_rule_d_by_d_ell_const_s}
\left ( \frac{\partial}{\partial \ell} \right )_s = \left ( \frac{\partial T}{\partial \ell} \right )_s \left ( \frac{\partial}{\partial T} \right )_P + \left ( \frac{\partial P}{\partial \ell} \right )_s \left ( \frac{\partial}{\partial P} \right )_T \, .
\end{equation}

\begin{figure}[!htp]

\includegraphics{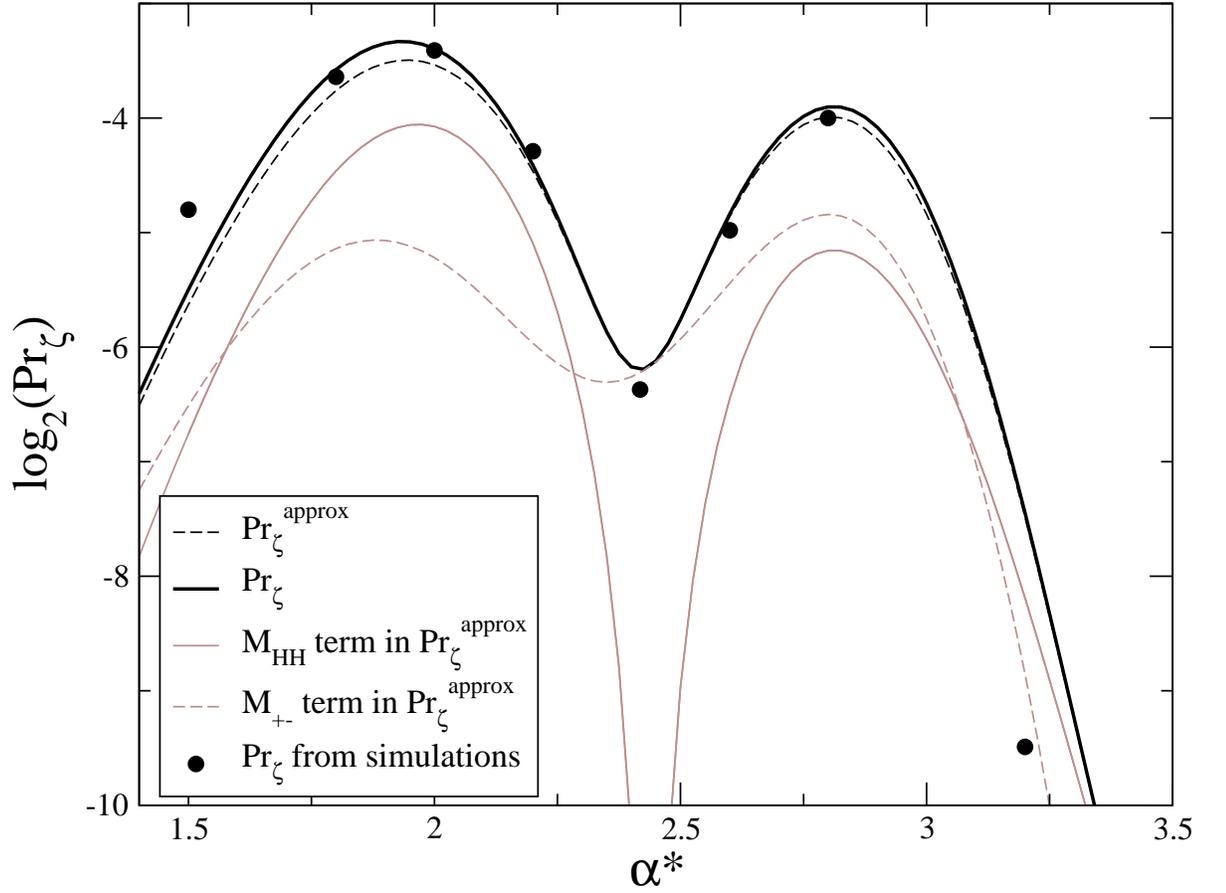}

\caption{\label{fig:Rvslambda}The value of the bulk Prandtl number, $Pr_{\zeta}$, as a function of the dimensionless parameter $\alpha^{\ast}$.  We have fixed $B=1$, $\beta = 1$, $m=1$, $P = 0$, and varied $\alpha^{\ast}$ by varying the cubic coefficient $\alpha$.  The simulation points for $\alpha^{\ast} =$ 1.5, and 3.2 are shown even though, at the lowest frequencies examined in those simulations, the hydrodynamic regime had not yet been reached.}

\end{figure}

\noindent The coefficient $(\partial P / \partial \ell)_s$ of the second term in (\ref{eq:chain_rule_d_by_d_ell_const_s}) is $-1/(\ell \chi_s) = -\gamma /(\ell \chi_T)$, which is expressed entirely in terms of quantities found in section~\ref{sec:cub-quart_chain}.  The coefficient of the first term can be rewritten

\begin{eqnarray}
\left ( \frac{\partial T}{\partial \ell} \right )_s & = & - \left ( \frac{\partial T}{\partial S} \right )_\ell \left ( \frac{\partial S}{\partial \ell} \right )_T =  - \frac{T}{c_V} \left ( \frac{\partial P}{\partial T} \right )_\ell = \frac{\gamma T}{c_P} \left ( \frac{\partial P}{\partial \ell} \right )_T \left ( \frac{\partial \ell}{\partial T} \right )_P \nonumber \\
& = & - \frac{\alpha_P \gamma T}{\chi_T c_P} \, ,
\end{eqnarray}

\noindent in which we have used the Maxwell relation $(\partial S /\partial \ell)_T = (\partial P / \partial T)_\ell$.  Thus, in summary, we can calculate $m_{+-}$ by

\begin{equation}
\frac{m_{+-}}{\ell} = \frac{\gamma}{\ell \chi_T} \left [ \frac{\ell \alpha_P T}{c_P} \left ( \frac{\partial \ln {(T\ell/ c)}}{\partial T} \right )_P + \left ( \frac{\partial \ln {(\ell/c)}}{\partial P} \right )_T \right ]
\end{equation}

\noindent which is expressed entirely in terms of the quantities found in section~\ref{sec:cub-quart_chain}.  The derivative calculations are very messy and algebraic packages such as Maple are very useful to ensure correctness of the final results.  With $m_{+-}$ and $m_{HH}$ in hand we can now proceed to calculate $Pr_{\zeta}$ for any value of $\alpha^{\ast}$ using (\ref{eq:Pr_soln}).  We thus obtain the theoretical curve shown in Fig.~\ref{fig:Rvslambda}.  As discussed earlier, we are restricting our attention to the region in parameter space surrounding $\alpha^{\ast} \simeq 2.418$.  At this value of $\alpha^{\ast}$ the value of $\gamma$ is maximized.  We see in Fig.~\ref{fig:Rvslambda} that $Pr_{\zeta}$ has a pronounced local minimum at $\alpha^{\ast} \simeq 2.418$.

The values of $Pr_{\zeta}$ obtained from simulations, described in the next section, can be seen in Fig.~\ref{fig:Rvslambda} as well.  The simulation results were obtained by varying $\alpha$ with $k_B T = 1$, $B = 1$, $P = 0$, $m = 1$.  As will be discussed, the simulations indicate that $\tilde{\hat{C}}_{\zeta} (\omega)$ and $\tilde{\hat{C}}_{\kappa} (\omega)$ go as the same power of $\omega$ as $\omega \to 0$ and so we find a well defined value of $Pr_{\zeta}$ at each value of $\alpha^{\ast}$.  As can be seen in Fig.~\ref{fig:Rvslambda}, the agreement between the simulations and the theory is good.  This supports the physical picture that we have been proposing.

\section{Numerical Results}

\begin{figure}[!htp]

\includegraphics[width=\textwidth]{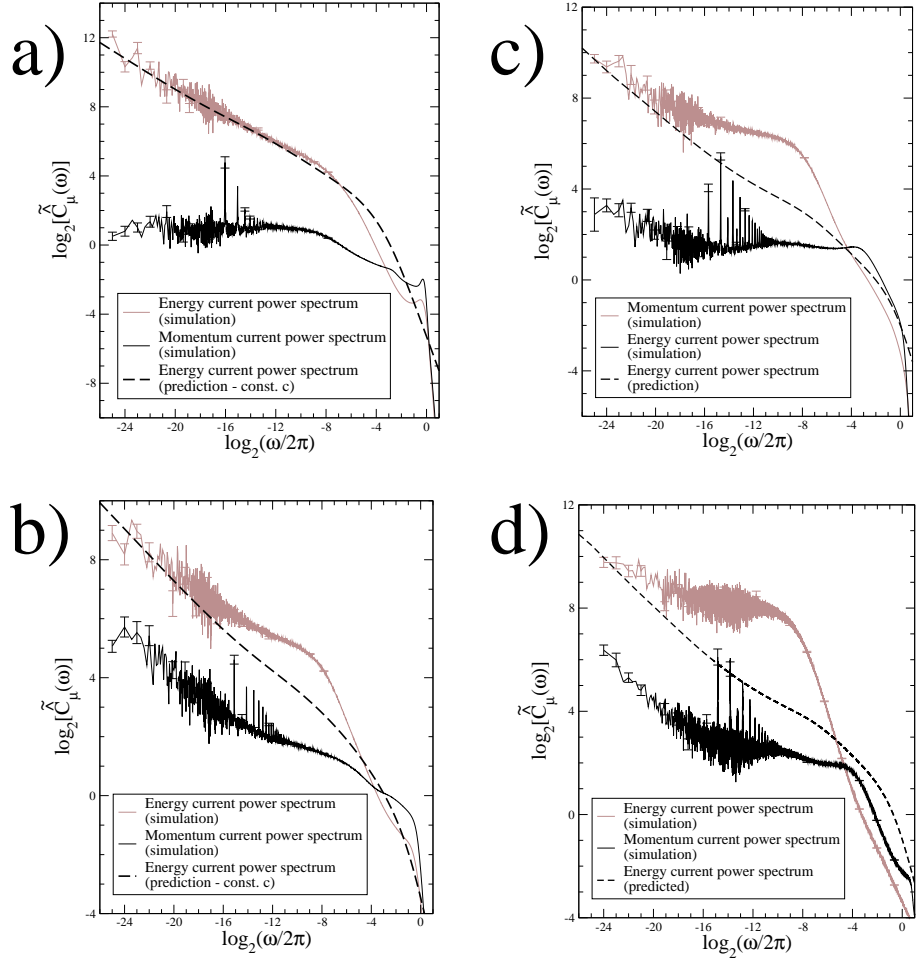}

\caption{\label{fig:all_ps}Current power spectra for a representative sample of values of $\alpha^{\ast}$.  a) $\alpha^{\ast} = 0$ (pure quartic),  b) $\alpha^{\ast}= 2.0$,  c) $\alpha^{\ast} = 2.418$, d) $\alpha^{\ast} = 2.8$.  The theoretically predicted curves for $\tilde{\hat{C}}_{\kappa}$ were obtained by the method described in \cite{pap:mine_PRE72}.  The approach of the observed energy power spectrum to the theoretical curve occurs at lower frequency as the value of $\alpha^{\ast}$ is increased.}

\end{figure}

We carry out molecular dynamics simulations of the system described by (\ref{eq:Hamiltonian}) in periodic boundary conditions for various values of the dimensionless parameter $\alpha^{\ast}$.  We choose $N = 2^{15}$ and typically run for $2^{25}$ time units, outputing $4$ times per time unit.  We set $m = 1$, $k_B T = 1$, $B = 1$ and $a = 0$.  We vary $\alpha^{\ast}$ by changing $\alpha$.  This, combined with our Hamiltonian (\ref{eq:Hamiltonian}), defines our units of length, time and energy.  In particular, distance and time are measured in units defined by (\ref{eq:length_time_units}).  Each run is initialized by randomly generating the positions and momenta of the particles according to an isobaric ensemble.  With the algorithms that we use we can achieve good accuracy using 8 time steps per time unit.  We output the sums over all particles of the momentum current and energy current.  Then, using the methods outlined in \cite{pap:mine_PRE72}, we calculate the momentum and energy current power spectra.  In this work we have used eighth order, sixth order and fourth order symplectic algorithms.  The eighth order algorithm was used for the $\alpha^{\ast} = 0$, $\alpha^{\ast} = 2.0$ and $\alpha^{\ast} = 2.418$ cases.  It is a refinement of the algorithms presented in \cite{Yoshida}.  This algorithm is the same as the one used in \cite{pap:mine_PRE72}.  It was realized later in this work that relaxing the symmetry used by Yoshida to obtain this eighth order algorithm allows the development of sixth order algorithms which are, nevertheless, more accurate than the eighth order one in terms of the error in energy.  Further, although the energy error was increased significantly by using a fourth order symplectic algorithm there was still no secular change in energy.  As a result, there is no appreciable loss in accuracy when a fourth order symplectic algorithm is used to find the total momentum and energy currents as we do in this work.  Hence, the later simulations in this work ($\alpha^{\ast} \geq 1.8$) were carried out using fourth or sixth order symplectic integrators.  These were checked against small numbers of runs using the eighth order routine to verify that the accuracy was preserved.  The coefficients of our fourth and sixth order integrators appear in Appendix A.

From our simulations we thus obtain $\tilde{\hat{C}}_{\zeta} (\omega)$ and $\tilde{\hat{C}}_{\epsilon} (\omega)$ (recall that in the Green-Kubo equation for $\hat{\kappa} (\omega)$ we can freely replace $\hat{C}_{\kappa}$ with $\hat{C}_{\epsilon}$).  We also use the theory developed in section II of \cite{pap:mine_PRE72} to produce a theoretical prediction of $\tilde{\hat{C}}_{\epsilon} (\omega)$.  Varying $\alpha^{\ast}$ we see that there appear to be distinct regimes in which the current power spectra behave in different ways.  Plots representative of these regimes are presented in Fig.~\ref{fig:all_ps}.  For reasons discussed in detail in \cite{pap:mine_PRE72} the $\alpha^{\ast} = 0$ case is special with $\tilde{\hat{C}}_{\zeta} (\omega) \to \textrm{const}$ and with $\tilde{\hat{C}}_{\epsilon} (\omega) \to \omega^{-1/2}$ as $\omega \to 0$.  For all nonzero values of $\alpha^{\ast}$, $\tilde{\hat{C}}_{\zeta} (\omega) \to \infty$ as $\omega \to 0$.  We note that the observed $\tilde{\hat{C}}_{\epsilon} (\omega)$ always approaches the theoretical prediction, but that this approach happens at progressively lower frequencies as $\alpha^{\ast}$ is increased.

Let us examine the different frequency regimes present in the transport in these systems.  In the highest frequency regime, defined by $\omega \tau >> 1$ where $\tau$ is a typical mode lifetime, damping plays no role \label{page:collisionless1}.  Accordingly we call this the collisionless regime.  This frequency regime will be examined in Appendix C.  At very small frequencies the hydrodynamic mode coupling model presented in \cite{pap:mine_PRE72} accurately predicts the energy current power spectrum.  This suggests ballistic transport of heat via low frequency sound modes which damp via a sound damping coefficient which is frequency dependent.  Further evidence that this is occuring in this regime is that, for $\alpha^{\ast} \neq 0$, $\tilde{\hat{C}}_{\zeta}$ and $\tilde{\hat{C}}_{\epsilon}$ go as the same power of $\omega$ in this regime.  This is consistent with the picture presented in our toy model presented in \cite{pap:mine_PRE72}.  This can be interpreted as evidence that local equilibrium is established and maintained by fast processes and gives rise to the mode coupling behaviour, as is assumed in \cite{Ernst2}.  We call this frequency regime the hydrodynamic regime.

\begin{figure}

\includegraphics[width=\textwidth]{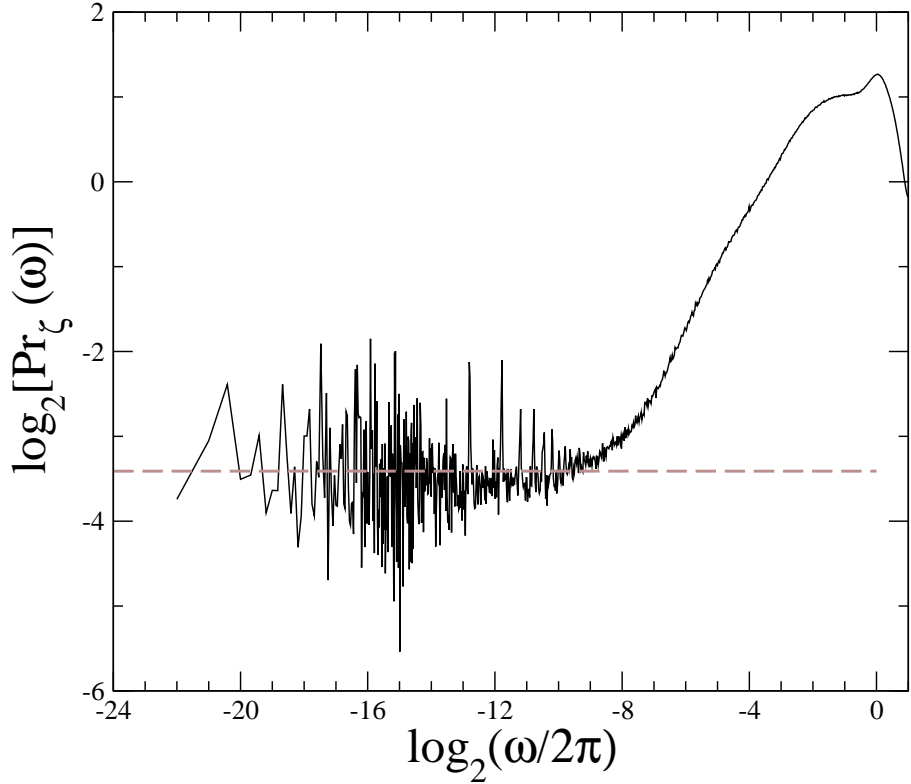}

\caption{\label{fig:Pr_vs_om} The frequency dependent bulk Prandtl number as obtained by simulation of the cubic-plus-quartic chain with $\alpha^{\ast} = 2.0$.  The horizontal dashed line is our asymptotic estimate and is given as a point in Figure~\ref{fig:Rvslambda}.}

\end{figure}

As we increase $\alpha^{\ast}$ we see that an intermediate frequency regime becomes established.  This regime has a $\tilde{\hat{C}}_{\epsilon} (\omega) \sim \omega^{-2}$ part which gives way to a constant $\tilde{\hat{C}}_{\epsilon} (\omega)$ plateau.  This is characteristic of the existence of a single relaxation time in this regime.  We note that as $\alpha^{\ast}$ increases the interparticle potential becomes more harmonic in the vicinity of its minimum.  Thus we might expect that, in some frequency regime, the behaviour of the system might look more and more like a harmonic lattice with small anharmonic perturbations as we increase $\alpha^{\ast}$.  This resembles the well known Boltzmann-Peierls picture of heat transport in which heat is carried at the sound velocity $c$ by phonons which scatter over some mean free path.  In the absence of defects the scattering is entirely due to phonon-phonon interactions.  It is known that, at least at first order, for 1D systems this damping cannot be due to the cubic term (three phonon processes) in the potential \cite{Lepri_review,pap:Peierls_AnnPhys_3}, so that the damping must be due to the quartic part of the potential (four phonon processes).  It is worth noting, however, that this argument is only valid if the cubic and quartic coefficients are small.  If the coefficients are as given in (\ref{eq:Vprime}) then the second order contribution due to three phonon processes can be comparable in size to the first order contribution due to four phonon processes.  In any case, given our speculation that the Boltzmann-Peierls picture is correct in the intermediate frequency regime for this system we refer to this intermediate frequency regime as the perturbative regime.  We give further evidence below that this picture is correct, though it should be stressed what we may be seeing is superposition or interference of second order cubic with first order quartic effects.  We discuss this regime still further in Appendix B where we work with a pure quartic model to avoid the difficulties of the second order cubic contributions.  A puzzle is that we see only a single relaxation time in the perturbative regime.

\begin{figure}

\includegraphics[width=\textwidth]{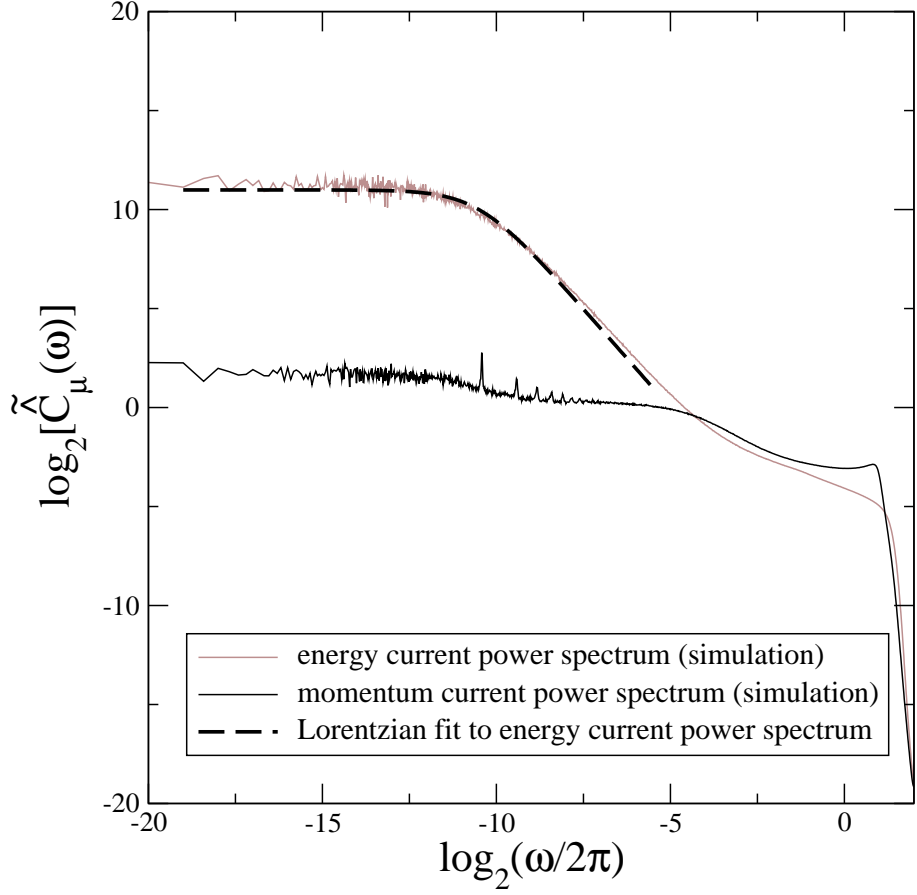}

\caption{\label{fig:alpha32}The current power spectra for the cubic-plus-quartic chain with $\alpha^{\ast} = 3.2$.  The hydrodynamic regime is at lower frequencies than are probed in this simulation.  The perturbative regime is very evident with the low frequencies following a Lorentzian shape.  A Lortentzian fit to the energy current power spectrum is shown.  The spectrum is well fit by $\tilde{\hat{C}}_\epsilon^{\textrm{fit}} = \tilde{\hat{C}}_\epsilon (0) /(1+\omega^2/\omega_0^2)$ with $\tilde{\hat{C}}_\epsilon (0) = 2.03 \times 10^3$, $\omega_0 = 6.91 \times 10^{-4}$.}

\end{figure}

It is simple to estimate $Pr_{\zeta} (\omega)$ from our simulation results by calculating $\tilde{\hat{C}}_{\zeta}/(m \beta \tilde{\hat{C}}_{\epsilon})$ for each frequency.  A typical example is shown in Fig.~\ref{fig:Pr_vs_om}.  At the lowest frequencies examined in the simulation $Pr_{\zeta} (\omega)$ seems to have converged to a constant value.  This is indicative of the hydrodynamic regime being reached at these frequencies.  For each value of $\alpha^{\ast}$ we take the average value of $Pr_{\zeta} (\omega)$ in the frequency range for which it appears to have converged and use this value as an estimate of $Pr_{\zeta}$.  The resulting values of $Pr_{\zeta}$ for the values of $\alpha^{\ast}$ examined are seen in Fig.~\ref{fig:Rvslambda}.  For values of $\alpha^{\ast}$ both higher and lower than the range examined, the frequency regime in which $Pr_{\zeta} (\omega)$ converges to a finite value is pushed to frequencies too low to be easily accessible.  The simulation points are shown for $\alpha^{\ast} = 1.5$ and $\alpha^{\ast} = 3.2$ even though the simulations did not probe low enough frequencies to see the hydrodynamic regime at those values of $\alpha^{\ast}$.  Thus the disagreement seen in Figure~\ref{fig:Rvslambda} for these two points should not be construed as a failure of the theory to predict $Pr_\zeta$.  Similar comments apply to simulations carried out for $\alpha^{\ast} =$ 3.6, 4.0, 4.2, 4.4 and 4.6.  Indeed, via methods from \cite{pap:mine_PRE72} we can predict the frequency at which we should see the crossover to the hydrodynamic regime and for $\alpha^{\ast} = 3.2$ the crossover should occur at $\omega/2\pi \simeq 2^{-26}$, so the lack of agreement seen for $\alpha^{\ast} \geq 3.2$ is expected.  For this reason we have not shown results for $\alpha^{\ast} > 3.2$ in Figure~\ref{fig:Rvslambda}.  In the range from $\alpha^{\ast} = 1.8$ to $\alpha^{\ast} = 2.8$, where the hydrodynamic regime was accessible, the agreement between simulation and the prediction from (\ref{eq:Pr_soln}) is good.  The standard errors in the simulation values are too small to be shown.  However, significant systematic errors are expected to be present since we are taking the average value of $Pr_{\zeta} (\omega)$ which is going asymptotically to a constant value.  In particular, the value of $Pr_{\zeta} (\omega)$ may not have been very close to its asymptotic value for the case of $\alpha^{\ast} = 2.418$.  The $Pr_{\zeta}$ vs. $\omega$ curve showed a discernable slope at the lowest frequencies examined for this value of $\alpha^{\ast}$.

Let us examine the perturbative regime where the behaviour resembles damping with a single relaxation time.  This regime covers a wider frequency range as we increase $\alpha^{\ast}$ beyond about 3.0.  An example of this is shown in Fig.~\ref{fig:alpha32} in which a Lorentzian fit is shown to $\tilde{\hat{C}}_\epsilon (\omega)$.  We must stress that this fit is not the $\omega \to 0$ limit of $\tilde{\hat{C}}_\epsilon (\omega)$, since the hydrodynamic regime is at lower frequencies than those shown in Fig.~\ref{fig:alpha32}.  If this behaviour applied as $\omega \to 0$ then we would see a finite $\kappa$.  Hence, if we ignore the behaviour in the lower frequency hydrodynamic regime, we may define an effective thermal conductivity in the single relaxation time regime which is just the zero frequency value that we obtain by a Lorentian fit to the heat current power spectrum in this regime.  This effective, perturbative regime, thermal conductivity is $\kappa_0^{\textrm{perturb}} = [\tilde{\hat{C}}_\epsilon^{\textrm{fit}} (\omega = 0)]/2$.  We can ask how this $\kappa$ depends on the temperature.  Noting that we can think of varying $\alpha^{\ast}$ as equivalent to varying $T$ we can define an effective temperature, $T^{\ast}$.  From (\ref{eq:dim_parameter_a}), if $B = 1$ and $\alpha = 1$, $\alpha^{\ast} \sim T^{-1/4}$.  Thus, we define

\begin{equation}
T^{\ast} = ({\alpha^{\ast}})^{-4} \, .
\end{equation}

\begin{figure}

\includegraphics[width=\textwidth]{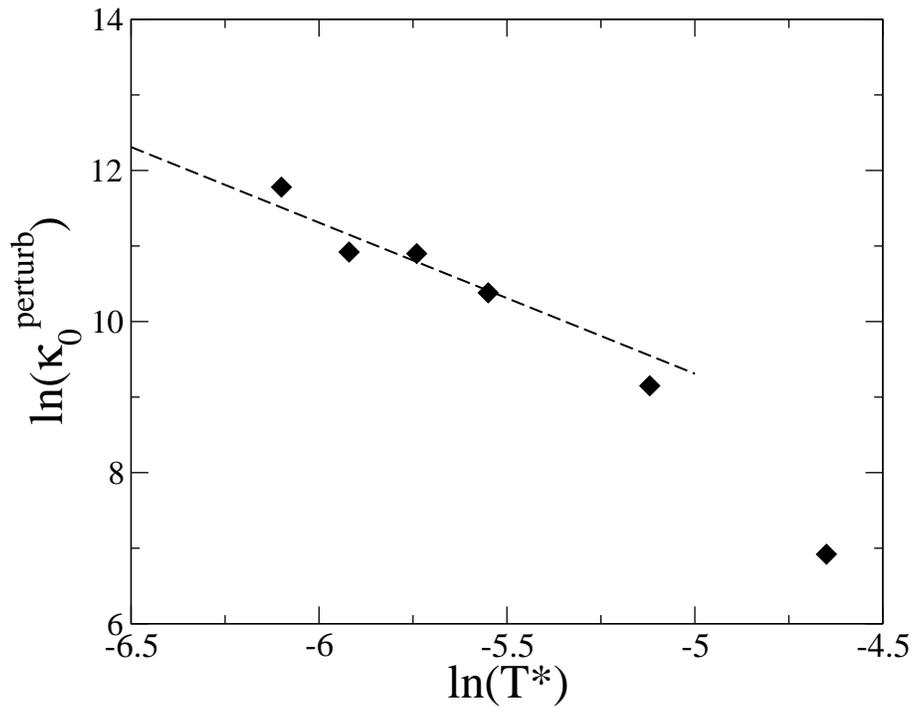}

\caption{\label{fig:k0vsTstar}The value of the perturbative regime plateau value of the thermal conductivity, $\kappa_0^{\textrm{perturb}}$, as a function of the effective temperature $T^{\ast}$ for low temperature (large $\alpha^{\ast}$).  We have fixed $B=1$, $\beta = 1$, $m=1$, and varied $\alpha^{\ast}$ by varying the cubic coefficient $\alpha$.  The dotted line is for reference and shows $(T^{\ast})^{-2}$ behaviour.}

\end{figure}

A plot of $\kappa_{0}^{\textrm{perturb}}$ vs. $T^{\ast}$, is shown in Fig.~\ref{fig:k0vsTstar} and we see that for low temperature $\kappa_{0}^{\textrm{perturb}}$ goes as $(T^{\ast})^{-2}$.  This is consistent with the well known result of a four phonon Boltzmann equation \cite{book:LifshitzPitaevskii,book:Peierls_QThofSolid}.  We see this as strong evidence that, for approximately harmonic lattices, there is a perturbative regime at frequencies higher than the hydrodynamic regime.  Whether this perturbative regime is well described by the four-phonon Boltzmann equation in every detail remains to be checked; a limited comparison is given in Appendix B.

In Appendix B we also indicate why a Boltzmann-Peierls equation approach fails to predict the chain bulk viscosity.

\section{Conclusions}

Several key points are worth pointing out immediately.

\begin{enumerate}
\item
For the cubic-plus-quartic system, in general, there are at least three distinct frequency regimes.  The lowest frequency regime, which we call the \emph{hydrodynamic regime}, is characterized by ballistic transport of heat via long wavelength sound waves.  The next regime, which we call the \emph{perturbative regime} resembles a harmonic chain with a quartic perturbation in some respects.  In this regime the transport of heat can be viewed as being damped by four phonon processes; the damping appears to be governed by a single relaxation time in the cubic-plus-quartic system but by a range of relaxation times in the absence of a cubic term as described in Appendix B.  At the highest frequencies observing times are too short for any appreciable phonon scattering.\label{page:collisionless2}  We call this the \emph{collisionless regime} and we examine it further in Appendix C.  A striking aspect of these results is that for nearly harmonic systems the perturbative regime and the hydrodynamic regime are clearly separated.  For more anharmonic systems the perturbative regime is not visible because the hydrodynamic regime is established at higher frequencies.

\item
The theory developed in \cite{pap:mine_PRE72} predicts the thermal conductivity well in the hydrodynamic regime.  This theory is based on an assumption of heat being ballistically carried by sound waves.  Thus, it is reasonable to conclude that the momentum transport and heat transport are strongly coupled.  This, combined with the fact that the power spectra have the same power law behaviour, is evidence that a theory like the toy model from the appendix of \cite{pap:mine_PRE72} describes the transport of both quantities. The current paper shows that, using the results of \cite{Ernst2}, we can extend the theory in \cite{pap:mine_PRE72} to predict the bulk viscosity as well. 

\item
The regime that we have called the perturbative regime is characterized by a $\hat{\kappa} \sim T^{-2}$ behaviour.  This is indicative of damping of modes by 4-phonon scattering (or likely a combination of first order effects from 4-phonon scattering and second order effects from 3-phonon scattering) as one would calculate using the four-phonon Boltzmann equation \cite{book:LifshitzPitaevskii,book:Peierls_QThofSolid}.  However, this identification is speculative.

\item
For a system like the one examined here with $\gamma \neq 1$ the bulk viscosity is infinite.  At sufficiently low frequencies \emph{the momentum current power spectrum has the same power law behaviour as the heat current power spectrum}.

\item
In the hydrodynamic regime the mode-coupling theory of Ernst et. al. \cite{Ernst1,Ernst2,Ernst3} fails to correctly predict the correct power law divergence of the transport coefficients.  However, it does predict the correct ratio between the thermal conductivity and the bulk viscosity (the bulk Prantdl number).  This theory assumes that local thermal equilibrium is maintained by fast processes whereas the hydrodynamic transport of heat and momentum is carried out by slower processes.  The success of the theory in predicting the bulk Prandtl number is evidence that the assumptions of this theory are correct for a 1D anharmonic chain with $\gamma \neq 1$.  Further, our results allow us to quantify what is meant by ``fast'' and ``slow'' in this system.  Specifically, the mode cascade of our model from \cite{pap:mine_PRE72} shows that on any time scale within the hydrodynamic regime the slow relaxation processes on that time scale are produced by much faster processes.  Thus, the meaning of ``slow'' is simply the time scale on which we are observing the relaxation process.  The meaning of ``fast'' is the frequencies for which $\Gamma_{k^{\prime}} {k^{\prime}}^2 \simeq \omega$ as is discussed following equation (10) in \cite{pap:mine_PRE72}.  Thus, the meanings of ``fast'' and ``slow'' vary with the time scale that we are examining.

\end{enumerate}

We have, in this paper and in \cite{pap:mine_PRE72}, described heat in 1D systems as being ballistically transported by sound waves which are weakly damped.  Purely ballistic transport would have no damping at all and would exhibit $\hat{C}_{\epsilon} (t) \sim t^{-1}$.  Ballistic transport of heat resembles second sound in which heat is transported at constant speed (the speed of ordinary sound in 1D); an important difference is that second sound exists only over very restricted temperature and frequency windows \cite{book:Ashcroft_Mermin} whereas ballistic transport is not subject to these restrictions.  It might seem to be more appropriate to refer to the transport seen here as superdiffusive transport ($\hat{C}_{\epsilon} (t) \sim t^{-p}$, $p < 1$).  However, we feel that the term ``superdiffusive transport'' implies a process similar to diffusion (a random walk where an initial distribution spreads with a speed which decreases with time) whereas what we are describing is much closer to ballistic transport in which a ``packet'' of heat is carried at constant speed over some long distance by a sound wave.  Further, if the toy model of \cite{pap:mine_PRE72} is correct then this produces a mode cascade with an infinite series of different power law behaviours of the transport coefficients.  This is quite distinct from the usual picture of superdiffusion.  Thus, we propose the term ``damped ballistic transport'' for this type of transport.

The existence of the perturbative and hydrodynamic regimes is highly significant when interpreting our results and the results of others.  It is possible for the plateau in the perturbative regime to look as if $\hat{\kappa}(\omega)$ is converging to a finite value at $\omega \to 0$.  However, if lower frequencies (larger systems) were examined the hydrodynamic regime would be seen and $\hat{\kappa}(\omega)$ would be seen to diverge.  Further, the extremely low frequency at which the hydrodynamic regime asserts itself in the system examined in this paper is below the lowest frequencies examined throughout much of the literature (for a summary of relevant results see \cite{Lepri_review}).  As always, extreme caution must be exercised in claiming that one is seeing the thermodynamic limit.

Another interesting feature of these results is that it seems to be valid to examine frequencies lower than the fundamental frequency of the system.  That is, we can observe processes on time scales longer than $\ell/c$, the time for a long wavelength disturbance to travel around the system one time.  Indeed, we can change the size of the system while holding our simulation times constant and see no change in the lowest frequency behaviour of the system even for frequencies below the fundamental frequency.  This is another consequence of the mode cascade.  For any system size we can observe relaxation processes to time-scales that the fundamental frequency dominates, and these slowest observed processes are much slower than the fundamental mode oscillations.  The lowest frequency transport modes that we are observing propagate around the system many times over the course of the simulation, but since their damping is driven by much faster processes it does not matter that they are circulating around a short chain rather than passing once through a very long chain.

There are several areas of further work which are needed to complete the above picture.  While the 4-phonon Boltzmann equation calculation has been done for the steady state case \cite{pap:Bernie}, no such calculation has been carried out for the equilibrium state.  This calculation should be carried out and the overall amplitude of the damping in the perturbative regime needs to be calculated so that the four-phonon Boltzmann equation can be used to quantitatively predict the heat current power spectrum in the perturbative regime.  Also, we do not have a satisfactory explanation for why, in the perturbative regime, a single relaxation time dominates.  This question might be answered by an equilibrium calculation.  A summary of the results of the nonequilibrium calculation in \cite{pap:Bernie} and analysis of how these results can be interpreted in the context of this study is presented in Appendix B.  Also discussed in Appendix B is the failure of the Boltzmann-Peierls approach for the calculation of the chain bulk viscosity.  Clearly a completely new approach is required for this property.

To show that the physical picture proposed in our toy model is correct, simulations would need to be carried out which probe systems to low enough frequencies to observe the onset of the next power law behaviour in the power spectra.  We predict that for the cubic-plus-quartic system at $\alpha^{\ast} = 2.0$ the ``kinks'' in the power spectra corresponding to the onset of the next power law behaviour in the series should occur somewhere below $\omega \sim 2^{-28}$.  For now, such low frequencies are beyond the practical limits of our simulations.  However, some other system might demonstrate the transitions between power laws at more accessible frequencies.

%\begin{acknowledgements}
%\end{acknowledgements}

\appendix

\section{Symplectic integrators}

There is by now a very large literature on improvements to the Yoshida integrators and a particularly extensive list of alternatives is given by \cite{pap:Omelyan_Comp_Phys_Comm151}.  We have not tested or used any of the versions that require calculations of the force gradient in addition to the force.  Instead we have opted for those schemes in which improvements are obtained solely by relaxing the requirement that the integrator contain the minimum possible number of steps.  This can reduce, in some cases dramatically, the size of the intermediate steps in the integrator and will typically improve both stability and accuracy at a small increase in running time.

	We have tested a limited number of integrators and give below two integrators, one 4th order and one 6th order, that we have found nearly optimal and used in production runs.  Both are ``position'' integrators which means the first move is a position update.  This is followed by a momentum update, then position, in the alternating sequence

\begin{eqnarray}
x_{new} & = & x + \frac{p}{m} w(1) \Delta t,  \quad p_{new} = p + f w(2) \Delta t \, ,\nonumber \\
x_{new} & = & x + \frac{p}{m} w(3) \Delta t,  \quad p_{new} = p + f w(4) \Delta t \ldots \, ,
\end{eqnarray}

\noindent where $f$ is the force.  The coefficients are as follows.

\begin{equation}
\begin{array}{lc}
\multicolumn{2}{c}{\textrm{4th order integration coefficients}} \\
w(1)=w(9) =  5/3/(3+\sqrt{39}) \, , & w(2)=w(8) =  3/4 \, , \\
w(3)=w(7) = -2/3/(6+\sqrt{39}) \, , &	w(4)=w(6) = -1/4 \, , \\
w(5) =  23/3/(4+\sqrt{39}) & \\
\end{array}
\end{equation}

\begin{equation}
\begin{array}{lcl}
\multicolumn{3}{c}{\textrm{6th order integration coefficients}} \\
w(1)=w(17) & =  & 0.055, \\
w(2)=w(16) & = & 0.1521292418198012208708832, \\
w(3)=w(15) & = & 0.2381129197090666122795175, \\
w(4)=w(14) & = & 0.3450759351638895426864652, \\
w(5)=w(13) & = & 0.5086956937118671097820404, \\
w(6)=w(12) & = & -0.0432317055841977505550037, \\
w(7)=w(11) & = & -0.4259353129298358880967277, \\
w(8)=w(10) & = & 0.0460265286005069869976553, \\
w(9) & = & 0.2482533990178043320703396 \\
\end{array}
\end{equation}

The 4th order coefficients are very close to the optimal equation (62) coefficients in \cite{pap:Omelyan_Comp_Phys_Comm151} but have the advantage, as analytical expressions, of being easily programmed for arbitrary precision.  The 6th order coefficients are based on the equation (83) ``momentum'' integrator of \cite{pap:Omelyan_Comp_Phys_Comm151} but have an added position step at the beginning which has been adjusted to further improve the integrator.  It is worth remarking that finding new solutions to the required non-linear constraint equations defining a symplectic integrator is a very difficult search problem but that modifying an existing one by small parameter increments is easy using Newton-Raphson iteration.

\section{Phonons and the Boltzmann-Peierls Equation}

	Our ultimate goal is to test the accuracy of the Boltzmann-Peierls approach to thermal transport in 1D chains in the frequency regime between the hydrodynamic regime discussed in the main sections of this paper and the collisionless regime discussed in Appendix C.  However, at the present time the theory is not well enough developed to make definitive tests possible.  In this appendix we report on the more modest achievement of a comparison of simulation with Boltzmann-Peierls in the relaxation time approximation.  Even this has only become possible because of the recent study \cite{pap:Bernie} by one of us (BN) of the Boltzmann-Peierls equation for a chain of weakly anharmonic oscillators in a thermal gradient.  An intermediate result in that study was an analytical formula that can be used to predict the wave-vector dependent relaxation rate of phonons in the relaxation time approximation.  This in turn allows us to predict the energy current power spectrum which we compare to simulations based on the model used in \cite{pap:Bernie}.  The results are good enough to unambiguously verify that the low frequency structure we see in the simulations is indeed due to phonon relaxation and that the Boltzmann-Peierls picture is qualitatively and even semi-quantitatively correct.

	The introductory part of our discussion relies heavily on results in \cite{pap:Bernie}.  To avoid excessive duplication we will use the same conventions as were adopted in that paper.  The relevant model is the FPU-$\beta$ model, namely equal mass particles described by the classical Hamiltonian similar to (\ref{eq:Hamiltonian})

\begin{eqnarray}
\label{eq:Bernie_H}
\mathcal{H} & = & \sum_i \mathcal{H} (p_i, X_i) \nonumber \\
& = & \sum_i \left ( \frac{1}{2}\frac{{p_i}^2}{m} + \frac{K_2}{2}{X_{i}}^2 + \frac{K_4}{4}{X_{i}}^4 \right ) \, ,
\end{eqnarray}

\noindent where the sum extends over the N particles in the chain with $X_{i+N}=X_i$ while $X_{i}$ is the deviation of a nearest neighbour pair separation from the equilibrium spacing $a$ and is defined as below (\ref{eq:Hamiltonian}).  In the weakly anharmonic limit we can use the harmonic part of $\mathcal{H}$ to define the normal modes of the system.  These are the phonons labelled by the $N$ independent wave-vectors $k_n=2\pi n/N$ and the phonon frequency

\begin{equation}
\label{eq:Bernie_dispersion_rel}
\omega_k = \omega_{ZB} \left | \sin{(\frac{k}{2})} \right |, \qquad \omega_{ZB} = 2\sqrt{\frac{K_2}{m}} 
\end{equation}

\noindent has its maximum $\omega_{ZB}$ at the Brillouin zone boundary while the phonon group velocity $u_k = d\omega_k /dk$ vanishes there.  The $K_4$ term in (\ref{eq:Bernie_H}) gives rise to phonon scattering and is treated in the Boltzmann-Peierls approximation in which the phonon occupation number $n_k$ is treated as a local density and its rate of change by transport, $\partial n_k/\partial t+ u_k \partial n_k /\partial x$, is equated to the rate of change by collisions, $R_k({n_{ki}})$.  In \cite{pap:Bernie} only the transport term $u_k \partial n_k /\partial x$ was relevant as the system was a steady state, non-equilibrium oscillator chain in a thermal gradient.  Here, for describing the relaxation of fluctuations from equilibrium in an otherwise spatially homogeneous chain, only $\partial n_k /\partial t$ is relevant.  As in \cite{pap:Bernie} we write the deviation from equilibrium in terms of a new function $g_k$ defined by

\begin{equation}
\delta n_k = n_k - \stackrel{eq}{\rule{0cm}{0.1cm}}\!\! n_k = \,\stackrel{eq}{\rule{0cm}{0.1cm}}\!\! n_k (\,\stackrel{eq}{\rule{0cm}{0.1cm}}\!\! n_k +1) g_k \, ,
\end{equation}

\noindent where $\stackrel{eq}{\rule{0cm}{0.1cm}}\!\! n_k$ is the Bose factor $1/(\exp(\hbar \omega_k/k_B T)-1)$.  The transport term in the Boltzmann-Peierls equation can then be written

\begin{equation}
\frac{\partial n_k}{\partial t} = \,\stackrel{eq}{\rule{0cm}{0.1cm}}\!\! n_k (\,\stackrel{eq}{\rule{0cm}{0.1cm}}\!\! n_k +1) \frac{\partial g_k}{\partial t} \, ,
\end{equation}

\noindent and this is to be equated to the net collision rate which to linear order in $g$ is

\begin{eqnarray}
\label{eq:Bernie_net_collision_rate}
R_k (g) & = & -\frac{9}{16 \pi}\hbar^2 \frac{{K_4}^2}{{K_2}^4}\int dk_1 \int dk_2 \omega_k \omega_{k_1} \omega_{k_2} \omega_{k_3} \nonumber \\
& & \times (g_k-g_{k_1}-g_{k_2}+g_{k_3}) \stackrel{eq}{\rule{0cm}{0.1cm}}\!\! n_k \,\stackrel{eq}{\rule{0cm}{0.1cm}}\!\! n_{k_3} \nonumber \\
& & \times (\,\stackrel{eq}{\rule{0cm}{0.1cm}}\!\! n_{k_1}+1)(\,\stackrel{eq}{\rule{0cm}{0.1cm}}\!\! n_{k_2} +1)  \delta (\omega_k-\omega_{k_1}-\omega_{k_2} + \omega_{k_3}) \, ,
\end{eqnarray}

\noindent where the $k_1$, $k_2$ integrations are understood to be over an interval of $2\pi$ and $k_3=k_1+k_2-k$ mod $2\pi$.  The rate (\ref{eq:Bernie_net_collision_rate}) is based on Fermi's golden rule and its derivation is standard textbook material that we need not repeat here.

The general solution to $\eqnk (\eqnk+1)\partial g_k /\partial t = R_k(g)$ can in principle be given in terms of the solutions to the associated eigenvalue equation but to our knowledge these have never been obtained and we do not attempt such solution here.  Instead we make what is known as the relaxation time approximation which is to set the deviations $g_{k_1}$, $g_{k_2}$, and $g_{k_3}$ in the integrand in (\ref{eq:Bernie_net_collision_rate}) to zero.  Then $R_k(g)$ becomes $g_k$ times an integral over equilibrium distributions which we can choose to write as the wave-vector dependent function $- \! \eqnk(\eqnk+1)/\tau_k$.  That is, with this approximation, the solutions to the Boltzmann-Peierls equation are

\begin{equation}
\label{eq:ind_relaxation_time}
g_k \propto \exp{(-t/\tau_k)} \, ,
\end{equation}

\noindent with

\begin{eqnarray}
\label{eq:Bernie_relax_rate_eqn}
\frac{\eqnk (\eqnk+1)}{\tau_k} & = & \frac{9}{16\pi}\hbar^2 \frac{{K_4}^2}{{K_2}^4} \int dk_1 \int dk_2 \omega_k \omega_{k_1} \omega_{k_2} \omega_{k_3} \nonumber  \\ 
& & \times \eqnk \,\stackrel{eq}{\rule{0cm}{0.1cm}}\!\! n_{k_3} (\,\stackrel{eq}{\rule{0cm}{0.1cm}}\!\! n_{k_1}+1)(\,\stackrel{eq}{\rule{0cm}{0.1cm}}\!\! n_{k_2} +1) \nonumber \\
& & \times \delta (\omega_k-\omega_{k_1}-\omega_{k_2} + \omega_{k_3}) \, .
\end{eqnarray}

\noindent Note that we are only interested in the classical limit in which case we can set $\eqnk \approx \eqnk+1 \approx k_BT/\hbar \omega_k$ and the relaxation rate equation (\ref{eq:Bernie_relax_rate_eqn}) becomes

\begin{subequations}
\begin{eqnarray}
\label{eq:relax_time}
\frac{1}{\tau_k} & = & \Omega \sin^2 \left (\frac{k}{2} \right ) K_k, \\
\Omega & = & \Omega (T) = \frac{9}{16\pi} \omega_{ZB} \left ( \frac{K_4 k_B T}{{K_2}^2} \right )^2, \\
K_k & = & \int dk_1 \int dk_2 \omega_{ZB} \delta (\omega_k-\omega_{k_1}-\omega_{k_2} + \omega_{k_3}) \, ,
\end{eqnarray}
\end{subequations}

\noindent where we have used also (\ref{eq:Bernie_dispersion_rel}) to express $\omega_k$ in terms of wave-vector $k$.

The dimensionless integral $K_k$ was evaluated in \cite{pap:Bernie} with the result

\begin{subequations}
\begin{eqnarray}
\label{eq:relax_time_integral}
K_k & = & 2 \left ( \frac{2}{3} \right )^{3/2} \left \{ z^{-1/6} B(1/3,1/3) F(1/3, 1/3; 2/3,;z)  \right . \nonumber \\
& & \left . - z^{1/6} B(2/3, 2/3) F(2/3, 2/3; 4/3; z) \right \} , \\
z & = & \frac{2}{27} \sin^2 \left ( \frac{k}{2} \right ) \, ,
\end{eqnarray}
\end{subequations}

\noindent where $B(x,y)=\Gamma(x)\Gamma(y)/\Gamma(x+y)$ is the beta function and the $F = \,_{2}\!F_{1}$ are hypergeometric functions.

The relaxation time approximation deserves some comment.  Were we to make the corresponding approximation of setting $g_{k_{1}} = g_{k_{2}} = g_{k_{3}} = 0$ in the thermal gradient problem in \cite{pap:Bernie}, we would find the amplitude of the leading divergence in the phonon distribution to be wrong by factor four.  However in that problem we know the distributions are driven by the thermal gradient and therefore the $g_k$ at different $k$ are strongly correlated.  In the equilibrium fluctuation problem we are considering here, it is quite reasonable to suppose that the $g_{k_i}$ ($k_i \neq k$) in (\ref{eq:Bernie_net_collision_rate}) are only weakly correlated with $g_k$ and thus on average nearly zero even when $g_k$ is not.  This makes the relaxation time approximation much less severe here; admittedly we cannot conclude anything quantitative.

The independent exponential relaxation in time of each normal mode as given in (\ref{eq:ind_relaxation_time}) implies the energy current power spectrum is a mode sum of the corresponding frequency domain lorenztians $\propto (1/\tau_k)/(\omega^2+(1/\tau_k)^2)$.  The explicit formula is

\begin{equation}
\label{eq:energy_curr_PS}
\tilde{\hat{C}}_{\epsilon} (\omega) = 2 (k_B T)^2 \int \frac{dk}{2\pi} {u_k}^2 \frac{1/\tau_k}{(\omega^2+(1/\tau_k)^2)} \, ,
\end{equation}

\noindent which also expresses the fact that energy in mode $k$ is transported at the group velocity $u_k = (1/2)\omega_{ZB}\cos(k/2)$.  The temperature dependent prefactor in (\ref{eq:energy_curr_PS}) can be deduced from the sum-rule requirement that $\int (d\omega/2\pi) \tilde{\hat{C}}_{\epsilon}(\omega)$ is the equal time energy current-current average $<\delta {\hat{j}_{\epsilon}}^2>$ which in the harmonic limit is $(k_BT)^2{\omega_{ZB}}^2/8$.  Given the explicit formulae (\ref{eq:relax_time}) for the relaxation time we can write $\tilde{\hat{C}}_\epsilon(\omega)$ in the scaling form

\begin{equation}
\label{eq:EcurrentPS_scaling_form}
\tilde{\hat{C}}_{\epsilon}(\omega) = (k_B T)^2 \frac{{\omega_{ZB}}^2}{2\Omega} \tilde{\hat{C}}^{R}(\omega/\Omega) \, ,
\end{equation}

\noindent where $\tilde{\hat{C}}^R(\omega)$ is the dimensionless spectrum

\begin{equation}
\label{eq:dimensionless_spectrum}
\tilde{\hat{C}}^R (\omega) = \int \frac{dk}{2\pi} \cos^2 \left ( \frac{k}{2} \right ) \sin^2 \left (\frac{k}{2} \right ) \frac{K_k}{\omega^2+\sin^4\left (\frac{k}{2}\right ){K_k}^2} .
\end{equation}

\noindent The integral in (\ref{eq:dimensionless_spectrum}) must be done numerically but then, as given by (\ref{eq:EcurrentPS_scaling_form}), can be applied universally with only an amplitude and frequency rescaling.  The spectrum $\tilde{\hat{C}}^R (\omega)$ is even in $\omega$, has the normalization

\begin{figure}

\includegraphics[width=0.8\textwidth]{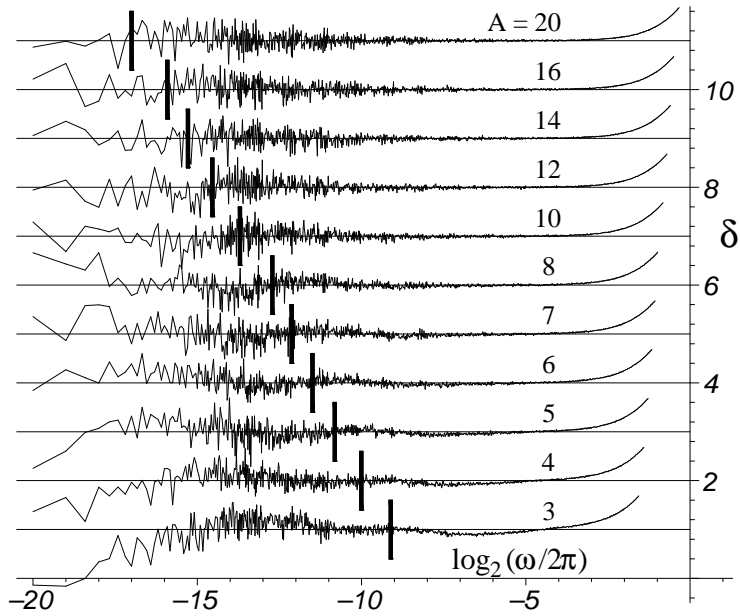}

\caption{\label{fig:deviation_sim_th} Energy current power spectrum deviation from best fit in the form $\delta = \log_2[\tilde{\hat{C}}_{\epsilon}(\omega)^{\textrm{Simulation}}/\tilde{\hat{C}}_{\epsilon}(\omega)^{\textrm{Eqn(\ref{eq:Ecurr_corr_fncn_final_form})}}]$ vs. $\log_2(\omega/2\pi)$.  The deviations have been shifted for clarity.  The relaxation rate correction factors for the curves from low to high $A$ are $R_{\tau} =$ 0.607, 0.893, 1.114, 1.317, 1.455, 1.547, 1.704, 1.803, 1.847, 1.910, 1.960.  The fit spectrum (\ref{eq:Ecurr_corr_fncn_final_form}) is characterized by the appropriately scaled asymptotes (\ref{eq:asymptotics1}) and (\ref{eq:asymptotics2}); their intersection defines a corner frequency $\omega_{\textrm{corner}} = 2.4443 R_{\tau} \Omega = R_{\tau} 2.4443 (9/8\pi)A^{-7/2}$ that is shown as the heavy vertical lines.}

\end{figure}

\begin{equation}
\int d\omega \tilde{\hat{C}}^R (\omega) = \frac{\pi}{2} \, ,
\end{equation}

\noindent and is characterized by simple power-law behaviour in the two limits of low and high frequency.  These are, for positive $\omega$,

\begin{subequations}
\label{eq:asymptotics}
\begin{eqnarray}
\label{eq:asymptotics1}
\tilde{\hat{C}}^R (\omega) & \approx & \omega^{-2/5} 0.2749172...\, ,\: (\omega \to 0) \\
\label{eq:asymptotics2}
& \approx & \omega^{-2} 1.1488360 \, ,\: (\omega \to \infty).
\end{eqnarray}
\end{subequations}

\noindent The $\omega^{-2/5}$ low frequency behaviour arises because of the small $k$ singularity in the relaxation time integral $K_k$ given in (\ref{eq:relax_time_integral}).  This in turn, as discussed in \cite{pap:Bernie}, is a consequence of the fact that the phonon dispersion curve $\omega_k$ at small $k$ is linear with a cubic correction.  Thus the $\omega^{-2/5}$ divergence observed here is quite general and not at all special to the nearest neighbour model (\ref{eq:Bernie_H}).  On the other hand, it may require the absence of odd terms in the potential.  The evidence for this is the frequency independent regime seen in the cubic-plus-quartic model spectrum in Fig.~\ref{fig:alpha32}.

For purposes of comparing the above theoretical results to numerical simulation we first note that we can rescale lengths and times such that the Boltzmann factor $\exp(-\mathcal{H}/k_B T)$ is reduced to a one parameter family.  The specific scaling we have chosen turns $\mathcal{H}(p, X)$ in (\ref{eq:Bernie_H}) into

\begin{subequations}
\begin{eqnarray}
\label{eq:H_rescaled}
\frac{\mathcal{H}(p, X)}{k_B T} \equiv \mathcal{H}(v, x) & = & \frac{v^2}{2} + A \half {x}^2 + \quarter {x}^4 \, ,\\
\label{eq:Adefn}
A & = & \frac{K_2}{(K_4 k_B T)^{1/2}} \, , 
\end{eqnarray}
\end{subequations}

\noindent and is equivalent to setting $m = K_4 = k_B T = 1, K_2 = A$.  Comparison with (\ref{eq:Vprime}) shows that, except for the cubic term, the FPU-$\beta$ model here is the (x-shifted) FPU-$\alpha \beta$ model of the text with $A=(\alpha^{\ast})^2$.  The case $A=0$ is the pure quartic model discussed in \cite{pap:mine_PRE72} whereas weak anharmonicity requires $A\to \infty$.  The largest $A$ in the simulations described below is $A=20$.

\begin{figure}

\includegraphics[width=0.8\textwidth]{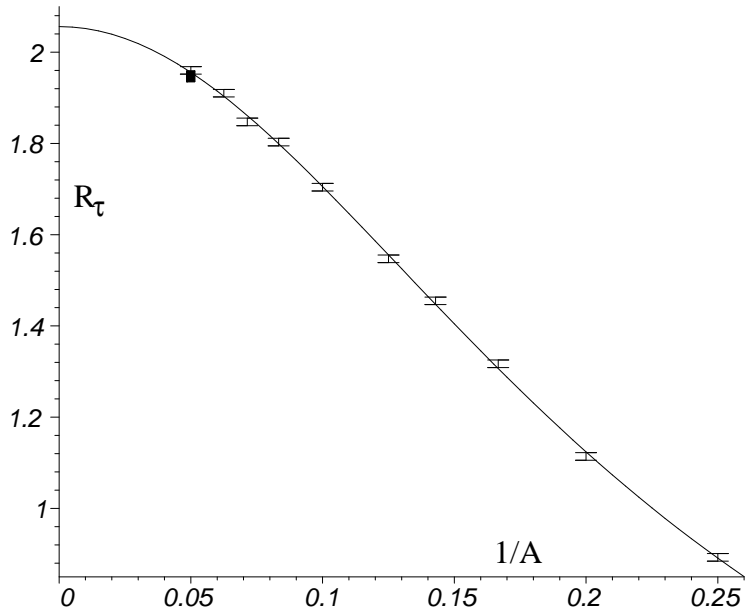}

\caption{\label{fig:relax_rate_correction} Individual $R_{\tau}$ estimates from Fig~\ref{fig:deviation_sim_th} vs. $1/A$.  The solid box near the top of the graph is the data from runs on chains of length $2^{15}$;  all other data points are for $N= 2^{13}$.  The data is well summarized for $A>3$ by the smooth curve which is $R_\tau = (2.056-0.10/A^2 - 4.66/A^4)/(1+20.47/A^2)$ and provides an estimate of the weak coupling limit $A\to \infty$ from our runs at finite $A$.  We use units with $K_4 (=B) = 1$, $\beta = 1$ and $m = 1$.}

\end{figure}

A number of thermodynamic properties of the model can be given explicitly in terms of known functions and are useful as numerical checks.  First note that because the potential terms in (\ref{eq:H_rescaled}) are even in $x$, the specific heat ratio $\gamma=1$.  We can then get the adiabatic sound speed, $c$, trivially from the isothermal compressibility and find

\begin{subequations}
\label{eq:therm_sound_speed_BP}
\begin{eqnarray}
\frac{c^2}{{c_0}^2} & = &\frac{2A^{-2}}{R_K-1} \, , \\
\label{eq:RK}
R_K & = & \frac{K_{3/4} (A^2/8)}{K_{1/4} (A^2/8)} \, ,
\end{eqnarray}
\end{subequations}

\noindent where $K_{3/4}$ and $K_{1/4}$ are Bessel functions and ${c_0}^2=A$ is the result one would have in the absence of the $x^4$ term in (\ref{eq:H_rescaled}).  Similarly, for the equal time energy current-current average we get

\begin{equation}
\label{eq:eq_time_E_current}
^{j} R_{j0} = <{\delta \hat{j}_{\epsilon}}^2>/<{\delta \hat{j}_{\epsilon}}^2>_0 = \frac{1}{2} \left [ 3 R_K - 1 \right ]
\end{equation}

\noindent where $<\delta {\hat{j}_{\epsilon}}^2>_0 = A/2$ is the result with no $x^4$ term in (\ref{eq:H_rescaled}) and equals $\int (d\omega/2\pi) \tilde{\hat{C}}_{\epsilon}(\omega)$ in the relaxation time approximation.  Thus any deviation of the ratio $^{j}R_{j0}$ in (\ref{eq:eq_time_E_current}) from unity indicates a failure of the power spectrum sum-rule.  This failure vanishes as $A\to \infty$ but is about 3\% at $A=10$ and rises to 11\% at $A=5$ and 27\% at $A=3$.  To some extent the failure is spurious and just a consequence of our having chosen to determine the amplitude in (\ref{eq:energy_curr_PS}) by evaluating averages in the harmonic limit.  A perfectly reasonable expedient which we adopt is to redefine $\tilde{\hat{C}}_{\epsilon}(\omega) = 2A \tilde{\hat{C}}^{R}(\omega/\Omega)/\Omega$ as given in (\ref{eq:EcurrentPS_scaling_form}) to $\tilde{\hat{C}}_{\epsilon}(\omega) = ^{j}\!\! R_{j0} 2A \tilde{\hat{C}}^{R}(\omega/\Omega)/\Omega$ and this ensures the sum-rule is exactly satisfied.  We also recognize that the relaxation time approximation for the mode decay rate, namely $1/\tau_k = \Omega \sin^2(k/2)K_k$ from (\ref{eq:relax_time}), is unlikely to be exactly correct.  The very simplest correction one can make here is a single multiplicative constant for all $1/\tau_k$ and hence we make the replacement $\Omega \to R_{\tau}\Omega$.  In conclusion, we compare

\begin{equation}
\label{eq:Ecurr_corr_fncn_final_form}
\tilde{\hat{C}}_{\epsilon} (\omega) =\; ^{j}\!R_{j0} \frac{2A \tilde{\hat{C}}^{R} (\omega/(R_{\tau} \Omega))}{R_{\tau}\Omega}
\end{equation}

\noindent to the simulation results with $R_{\tau}$ as adjustable and designated as the relaxation rate correction factor.

Comparison with simulation is shown in Fig.~\ref{fig:deviation_sim_th} in the form of a logarithmic deviation, namely $\log_2[\tilde{\hat{C}}_{\epsilon}(\omega)^{\textrm{Simulation}}/\tilde{\hat{C}}_{\epsilon}(\omega)^{\textrm{Eqn(\ref{eq:Ecurr_corr_fncn_final_form})}}]$ vs. $\log_2(\omega/2\pi)$.  Values for $A$ range from 3 to 20 and it is only for the smallest $A$ that any significant systematic deviation in the low frequency phonon region can be detected.  Each fit in Fig.~\ref{fig:deviation_sim_th} has yielded a relaxation rate correction factor and these are shown in Fig.~\ref{fig:relax_rate_correction} as $R_{\tau}$ vs. $1/A$.  Any variation with $A$ indicates that there are processes contributing to the scattering beyond that given by the Fermi golden rule result (\ref{eq:Bernie_relax_rate_eqn}) and the data in Fig.~\ref{fig:relax_rate_correction} is consistent with these perturbative corrections scaling asymptotically as $1/A^2$.  This is expected since, as given in (\ref{eq:Adefn}), $1/A$ is proportional to $\sqrt{K_4}$ in the original Hamiltonian and thus the corrections vary linearly with perturbation, $K_4$.  Most significantly, the simulation results show that there is a limiting value for $R_{\tau}$ as $A\to \infty$ and, thus, are consistent with the scaling expected from the Boltzmann-Peierls analysis.  The fact that this limiting value for $R_{\tau}$ is not unity but $R_{\tau} \approx 2.0-2.1$ means our analysis is not yet truly quantitative.  We have no way of knowing whether the observed rate correction factor is the result of the relaxation time approximation, failure of the Boltzmann ``Stosszahlansatz'' in the context of 1D phonons, or some combination of the two.  Resolution will have to wait until further theoretical work is completed.

The above discussion suggests that we could calculate the corresponding chain viscosity from the Boltzmann-Peierls equation.  We indicate briefly why this approach fails.  The physical picture underlying the calculation of viscous dissipation in (3D) insulating solids is due to Akhieser \cite{pap:akhieser_JPhysUSSR1} -- see also \cite{pap:DeVault_PhysRev155,book:Gurevich,pap:Woodruff_PhysRev123}.  Imagine one of the low frequency phonons in a lattice system of phonons.  What is meant by low frequency in this case is $\omega \tau < 1$, where $\tau$ is the average relaxation time (due to phonon collisions) of all phonons in the system.  The given phonon can be thought of as slowly modulating the spatial density.  Therefore, because of anharmonicity in the interparticle interactions, the elastic constants and frequencies of all phonons are also modulated by the phonon in question.  The mode Gruneisen parameter $\gamma_k$, where for 1D \cite{Ashcroft_Mermin_pg493}

\begin{equation}
\gamma_k = -\frac{\langle L \rangle}{\omega_k} \frac{\partial \omega_k}{\partial \langle L \rangle}=-\frac{\partial (\ln{\omega_k})}{\partial (\ln{\langle L \rangle)})} \, ,
\end{equation}

\noindent describes the frequency shift of mode $k$ due to the lattice strain induced by the given phonon.  In general (3D) the $\gamma_k$ vary from mode to mode, so that the differing frequency shifts put the system temporarily out of thermal equilibrium (recall that the thermal mode occupancies $n_k$ are adiabatic invariants and thus temporarily constant -- i.e.until relaxation occurs -- for small $\omega$, and that the new $n_k$ will depend on the shifted frequencies).  If $\omega$ is small enough (i.e. $\omega \tau \ll 1$) the system can relax back to equilibrium during the period of modulation, $2\pi/\omega$, of the lattice and the distortion process is quasi-static (i.e. reversible or dissipationless).  For larger $\omega$, say $\omega \tau \lesssim 1$, the relaxation will not precisely follow the distortion, leading to irreversible behaviour with viscous dissipation.

For frequencies $\omega \tau \lesssim 1$ Akhieser and others \cite{pap:akhieser_JPhysUSSR1,pap:DeVault_PhysRev155,book:Gurevich,pap:Woodruff_PhysRev123} have applied the Boltzmann-Peierls equation, with and without the single relaxation time approximation, to calculate 3D lattice viscosities (i.e. shear and bulk).  If we apply this theory to 1D chains we find that the bulk viscosity is a sum of contributions from all modes, and that the contribution of mode $k$ contains the factor $(\delta \gamma_k)^2$ ,where $\delta \gamma_k = \gamma_k - \langle \gamma \rangle$, with $\langle \ldots \rangle$ denoting an average over all $k$.  As we shall see, for our purposes the exact nature of the averaging over $k$ need not be specified.  The appearance of the quantity $\delta \gamma_k$ is unsurprising in view of the physical picture given above.  We now evaluate $\delta \gamma_k$ for a 1D chain with Hamiltonian of type (\ref{eq:Bernie_H}) and (\ref{eq:Hamiltonian}), with $K_2$, $K_3$, and $K_4$ denoting the harmonic, cubic and quartic spring constants, respectively.  The value of $\gamma_k$ for such a chain is \cite{Ashcroft_Mermin_pg508}

\begin{equation}
\gamma_k = \left ( \frac{a}{2} \right ) \left ( \frac{K_3}{K_2} \right ) \, .
\end{equation}

\noindent Note that $\gamma_k$ is independent of $k$, so that $\delta \gamma_k = 0$.  Thus the Boltzmann-Peierls approach fails to predict a viscosity for all such chains (with nearest-neighbour interactions), for arbitrary values of $K_3$ and $K_4$.

\section{Momentum current in the collisionless regime}

We verify statements made on pages \pageref{page:collisionless1} and \pageref{page:collisionless2} of the text that, at least for a specific model, the high frequency weak coupling regime is a distinct collisionless regime in which phonon scattering plays no role.  Explicitly, we show that the momentum current correlation function for the FPU-$\beta$ model of Appendix B, evaluated as an average over a thermal population of undamped phonons, agrees with numerical simulation.  Moreover, the comparison with the numerical simulation shows in what frequency regime other processes dominate the momentum current power spectrum.

The momentum current for the FPU-$\beta$ model of Appendix B is given by

\begin{equation}
\label{eq:tau}
\delta \hat{j}_\zeta = - \frac{K_2}{\sqrt{N}} \sum_i X_{i} - \frac{K_4}{\sqrt{N}} \sum_i {X_{i}}^3
\end{equation}

\noindent and it is a straightforward, albeit tedious, textbook exercise to express the coordinate deviations $X_{i}$ in terms of phonon creation and annihilation operators and evaluate the average $\langle \delta \hat{j}_\zeta (t) \delta \hat{j}_{\zeta} (0) \rangle$ in an ensemble of non-interacting, Bose distributed phonons.  The terms proportional to ${K_2}^2$ and $K_2K_4$ are time independent because the coordinate sums in (\ref{eq:tau}) dictate that there are no remaining $k \neq 0$ phonons in the final expressions.  The term proportional to ${K_4}^2$ is a three phonon average in which the total momentum $k_1+k_2+k_3 = 0$.  The final expression for the temporal Fourier transform of the correlation function, in the high temperature or classical limit, is

\begin{equation}
\label{eq:FT_corfnc}
\tilde{\hat{C}}_{\zeta} (\omega) = \frac{3}{\pi} \left ( \frac{2 k_B T}{m {\omega_{ZB}}^2} \right )^3 {K_4}^2 \int dk_1 \int dk_2 \sum \delta (\pm \omega_1 \pm \omega_2 \pm \omega_3-\omega).
\end{equation}

\noindent The integrals in (\ref{eq:FT_corfnc}) are understood to be over an interval of $2\pi$ and the remaining sum is meant to indicate eight separate terms corresponding to all possible sign combinations in the $\delta$-function.  All of these terms can be combined into one by the expedient of a factor two, dropping the absolute value signs in the definition (\ref{eq:Bernie_dispersion_rel}) of the phonon frequency, and extending the integration interval for each momentum to $4\pi$.  The result is our collisionless power spectrum

\begin{equation}
\label{eq:collisionless_spect}
\tilde{\hat{C}}_{\zeta}(\omega) = \frac{3}{\pi} \frac{k_B TK_2}{\omega_{ZB}} A^{-4} I(\omega/\omega_{ZB})
\end{equation}

\noindent with $A$ given by (\ref{eq:Adefn}) and $I(z)$ the dimensionless spectrum

\begin{equation}
\label{eq:dimless_spect}
I(z) = 1/4 \int dk_1 \int dk_2 \delta \left [\sin \left(\frac{k_1}{2} \right)+\sin \left ( \frac{k_2}{2} \right )-\sin \left ( \frac{k_1}{2}+ \frac{k_2}{2} \right ) -z \right ]
\end{equation}

\noindent in which integration intervals of $4\pi$ are understood making $I(z)$ a full period integral over a periodic function.

The evaluation of $I(z)$ will be given below but a number of important features can be understood without detailed calculation.  A key observation is that (\ref{eq:dimless_spect}) is completely analogous to a density of states expression commonly found in solid state physics and as such has singular points which are the well known van Hove singularities.  The three sine function sum in the $\delta$-function has several symmetry related local maxima which are also the absolute maxima.  One of these is at $k_1=k_2=4\pi/3$ and this implies the spectrum approaches a constant as $z$ approaches $3\sqrt{3}/2$ from below and vanishes identically for $z>3\sqrt{3}/2$.  Since $I(z)$ is an even function of $z$, there is a corresponding cutoff at $z=-3\sqrt{3}/2$.  The integrand region near the origin, $k_1=k_2=0$, is also the source for a van Hove singularity and this is rather more unusual and interesting.  Because the linear terms in the sine sum vanish, the leading behaviour is the cubic proportional to $k_1 k_2 (k_1+k_2)$ and simple power counting then shows the spectrum must diverge as $|z|^{-1/3}$.  We can expect this to be a general result, dependent only on the fact that phonon dispersion curves vary linearly with cubic corrections at small $k$, and not specific to the nearest neighbour force constant model being treated here.

The evaluation of (\ref{eq:dimless_spect}) is quite involved with a lot of similarities to what was done in \cite{pap:Bernie} to obtain the relaxation time integral $K_k$ that has been reproduced here as (\ref{eq:relax_time_integral}).  The first step is the change of variables $k_1=2(u+v)$, $k_2=2(u-v)$ which transforms the sine sum into $2\sin(u) \cos(v)- \sin(2u)$ and makes the $v$ integration trivial.  The remaining $u$ integral is

\begin{equation}
\label{eq:u_int}
I(z) = 4\int \frac{du}{\sqrt{4\sin^2 (u)-(\sin (2u)+z)^2}}
\end{equation}

\noindent with the $u$ integration over that interval between 0 and $\pi$ for which the argument of the square root is non-negative.  The substitution $x=\cot (u)$ puts the integrand in (\ref{eq:u_int}) into algebraic form and shows the result can be expressed as an elliptic integral.  The most convenient result however is obtained by expanding the elliptic integral as a series and recognizing the series as the expansion of

\begin{eqnarray}
\label{eq:elliptic_expansion}
I(z) & = & \left ( \frac{4}{z} \right )^{1/3} B(1/3,1/3) F(1/3,1/3;2/3;4z^2/27) \nonumber \\
& & - \frac{4}{3} \left ( \frac{z}{4} \right )^{1/3} B(2/3,2/3) F(2/3,2/3;4/3;4z^2/27)
\end{eqnarray}

\noindent which, surprisingly, except for normalization and a change of argument $4z^2/27 \leftrightarrow 2\sin^2(k/2)/27$, is the $K_k$ integral (\ref{eq:relax_time_integral}).  The final proof that (\ref{eq:elliptic_expansion}) is the correct value of (\ref{eq:u_int}) is completed by showing both forms satisfy the same differential equation.  All of the above manipulations are complicated and could not have been done without computer packages such as Maple.

\begin{figure}

\includegraphics{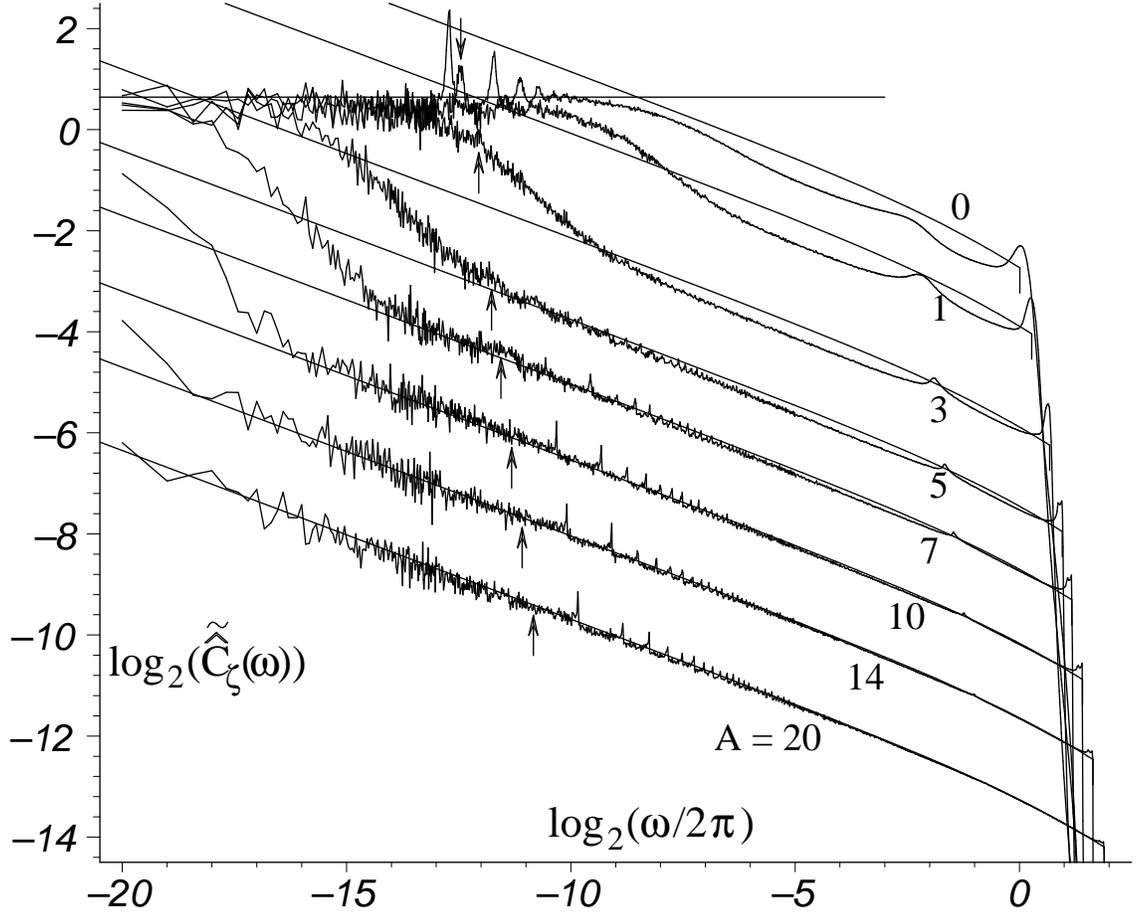}
\caption{\label{fig:MCPS_collisionless}Comparison of simulations of the momentum current power spectrum with the collisionless spectrum (\ref{eq:collisionless_spect}) and (\ref{eq:elliptic_expansion}) rescaled as described in the text.  The horizontal line is the estimated amplitude for the zero frequency limit of the spectrum for the pure quartic model taken from \cite{pap:mine_PRE72} and is based on simulations on chains longer than those treated here.  Arrows on curves for $A>0$ mark the expected frequency for the longest possible wavelength mode in the system.  Even harmonics of this fundamental are apparent in the large $A$, weakly anharmonic, simulations; the fundamental itself is not visible except at small $A$.  As in Fig.~\ref{fig:deviation_sim_th}, the simulations are run with $K_4 = \beta = m = 1$, $K_2 = A$.}
\end{figure}

The power spectrum (\ref{eq:collisionless_spect}) with the explicit result (\ref{eq:elliptic_expansion}) has been compared with the numerical simulations described in Appendix B.  The agreement is excellent for the largest $A$ values and can be improved for smaller $A$ by two renormalizations similar to what was done for the energy current power spectrum in Appendix B.  First, the harmonic model zone boundary frequency $\omega_{ZB}=2\sqrt{K_2/m}$ is rescaled by the sound velocity ratio $c/c_0$ using (\ref{eq:therm_sound_speed_BP}).  This guarantees that $\omega=\omega (k)$ in the limit $k \to 0$ is given exactly even when, as $A$ is decreased, the modes at large $k$ strongly damp and their frequency becomes ambiguous.  Second, the spectrum amplitude is rescaled to give the exact frequency sum-rule.  The integral over the collisionless $\tilde{\hat{C}}_{\zeta}(\omega)$ given in (\ref{eq:collisionless_spect}) is

\begin{equation}
\label{eq:colnless_sumrule}
\int d\omega \tilde{\hat{C}}_{\zeta}(\omega) = \frac{12 \pi k_B T K_2}{A^4}
\end{equation} 

\noindent whereas the exact result, as obtained from the equal time thermodynamic average $2\pi \langle \delta {\hat{j}_{\zeta}}^2 \rangle$, is

\begin{equation}
\int d\omega \tilde{\hat{C}}_{\zeta}(\omega) = \pi k_B T K_2 \left [ 3R_K-1-\frac{4}{R_K A^2} \right ] 
\end{equation}

\noindent where $R_K$ is the Bessel function ratio (\ref{eq:RK}).  The exact expression reduces, in the two limiting cases of large and small $A$ to

\begin{eqnarray}
\int d\omega \tilde{\hat{C}}_{\zeta}(\omega) & = & \frac{12\pi k_B T K_2}{A^4} \left [ 1-\frac{18}{A^2}+O(1/A^4) \right ] \nonumber \\
& = & \frac{\pi k_B T K_2}{A} \left [ 12R_{\Gamma}-\frac{1}{R_{\Gamma}}+O(A) \right ],  
\end{eqnarray}

\noindent where

\begin{equation}
R_{\Gamma}=\frac{\Gamma(3/4)}{\Gamma(1/4)}.
\end{equation}

\noindent Thus we confirm (\ref{eq:colnless_sumrule}) in the limit $A \to \infty$ and yet obtain a finite result for $A \to 0$.

Comparisons of the rescaled predictions with simulation are shown in Fig.~\ref{fig:MCPS_collisionless}.  Remarkably, even the $A=0$, pure quartic, model is in qualitative agreement at high frequencies with the collisionless approximation.  At lower frequencies we see another contribution to the simulation spectrum which presumably is related to the finite lifetime of the phonons.  Exactly what this relation might be is, however, not obvious since, as shown in Appendix B, the Boltzmann-Peierls approach will not yield a non-zero viscosity.  Also, the corner frequencies at which the spectrum saturates in the momentum current power spectrum are much lower than those in the energy current spectrum as seen by comparison of Figures \ref{fig:MCPS_collisionless} and \ref{fig:deviation_sim_th}.  Another striking observation from Fig.~\ref{fig:MCPS_collisionless} is that the low frequency saturation value of the momentum current spectrum is independent of the magnitude of the anharmonic term in the hamiltonian.  Such universality begs for an explanation; we do not have one.

\bibliographystyle{spmpsci}      % mathematics and physical sciences
\bibliography{refs}   % name your BibTeX data base

%\begin{thebibliography}{}
% etc
%\end{thebibliography}

\end{document}